\documentstyle[12pt,epsfig,graphicx,amsfonts,psfrag]{article}

\makeatletter

\@addtoreset{equation}{section} \makeatother

\newcommand{\nn}{\nonumber}
\newcommand{\be}{\begin{equation}}
\newcommand{\ee}{\end{equation}}
\newcommand{\ba}{\begin{eqnarray}}
\newcommand{\ea}{\end{eqnarray}}

\newcommand{\dle}[1]{\label{#1}}
\newcommand{\dla}[1]{\label{#1}}
\newcommand{\dr}[1]{\ref{#1}}
\newcommand{\dc}[1]{\cite{#1}}
\newcommand{\dbibitem}[1]{\bibitem{#1}}
\newcommand{\gsim}{\raise.3ex\hbox{$>$\kern-.75em\lower1ex\hbox{$\sim$}}}
\newcommand{\lsim}{\raise.3ex\hbox{$<$\kern-.75em\lower1ex\hbox{$\sim$}}}

\newcommand{\matr}[1]{{\bf{#1}}}
\newcommand{\tr}{{\rm tr}}

\newcommand{\cA}{{\cal A}}
\newcommand{\cB}{{\cal B}}
\newcommand{\cC}{{\cal C}}
\newcommand{\cD}{{\cal D}}
\newcommand{\cE}{{\cal E}}

\newcommand{\cG}{{\cal G}}
\newcommand{\cL}{{\cal L}}
\newcommand{\cM}{{\cal M}}
\newcommand{\bgamma}{{\mbox{\boldmath $\gamma$}}}

\newcommand{\ha}{\mbox{$\textstyle\frac{1}{2}$}}

\newcommand{\qa}{\mbox{$\textstyle\frac{1}{4}$}}

\begin{document}

\renewcommand{\thefootnote}{\fnsymbol{footnote}}

\vskip 12pt

\begin{center}

{\large\bf Brane cosmology, varying speed of light and inflation in models
with one or more extra dimensions\footnote{Proceedings of the Peyresq-6
Meeting on ``Cosmological Inflation and Primordial Fluctuations.  Energy
desert and sub-millimeter gravity'', June 23-29, 2001.}\vskip 0.1cm}

\vskip 1.2cm

D.A.Steer$^{a}$\footnote{{\tt steer@th.u-psud.fr}} and
M.F.Parry$^{b}$\footnote{{\tt
parry@theorie.physik.uni-muenchen.de}}
\\
\vskip 5pt \vskip 3pt {\it a}) Laboratoire de Physique Th\'eorique,
B\^at 210, Universit\'e Paris XI, \\
Orsay Cedex, France\vskip 3pt {\it b})
Theoretische Physik,
Ludwig-Maximilians Universit\"{a}t,
Theresienstra\ss e 37, D-80333~M\"{u}nchen, Germany\\
\end{center}

\vskip 1.2cm

\renewcommand{\thefootnote}{\arabic{footnote}}
\setcounter{footnote}{0} \typeout{--- Main Text Start ---}

\begin{abstract}

We summarise the approach to brane cosmology known as ``mirage
cosmology'' and use it to determine the Friedmann equation on a
3-brane embedded in different bulk spacetimes all with one or more
extra dimensions.  Usually, when there is more than one extra
dimension the junction conditions, central to the usual brane
world scenarios, are difficult to apply. This problem does not
arise in mirage cosmology because the brane is treated as a ``test
particle'' in the background spacetime. We discuss in detail the
dynamics of a brane embedded in two specific 10D bulk spacetimes,
namely Sch-AdS$_5 \times$S$_5$ and a rotating black hole, and from
the dynamics---which are now rather more complicated since the
brane can move in all the extra dimensions---determine the new
``dark fluid'' terms in the brane Friedmann equation. Some of
these, such as the cosmological constant term, are seen to be bulk
dependent.  However, for both bulks we show that there exists a critical
brane angular momentum, $\ell_c$, and discuss its significance.
We then show explicitly how this mirage cosmology
approach matches with the familiar junction condition approach
when there is just one extra dimension. The issue of a varying
speed of light in mirage cosmology is reviewed and we find a
scenario in which $c_{\bf eff}$ always increases, tending
asymptotically to a constant $c_0$ as the universe expands.
Finally some comments are made regarding brane inflation and
limitations of the mirage cosmology approach are also discussed.

\end{abstract}

\section{Introduction}\dle{s:intro}

Recently there has been much interest in the idea that our
universe may be a 3-brane embedded a spacetime of five or more
dimensions.  In particular, following the work of Randall and
Sundrum \dc{RS}, the brane cosmology in models with one infinite
extra dimension has been studied in depth
\dc{Binetruy,JC1,JC2,JC3,Kraus,Ida}. There are essentially two
distinct approaches to determining this brane cosmology. In the
first (e.g.\ \dc{Binetruy}), coordinates are chosen relative to
the brane which is thus at a fixed position in the extra
dimension. The bulk 5D metric, on the other hand, is time
dependent and this time dependence induces a time dependence on
the brane via the junction conditions. The resulting Friedmann
equation on the brane is found to have a characteristic $\rho^2$
term \dc{Binetruy}, where $\rho$ is the energy density in matter
which is assumed to be confined on the brane, as well as a `dark'
radiation term originating from the Weyl tensor in the bulk
\dc{Weyl}.

In the alternative but equivalent approach (e.g.\ \dc{Kraus,Ida}),
the bulk is static and the brane dynamical: the brane moves
through a time-independent bulk metric. If the vacuum Einstein
equations hold in the bulk, and if one imposes that our universe
brane has the symmetry of a 3-sphere, then it is possible to prove
that the bulk must be Sch-AdS$_5$ \dc{Charm,CU}. Thus the brane
divides two regions of Sch-AdS$_5$ and its dynamics can be
determined from the junction conditions.  For reasons which will
be summarised in section \dr{s:mirage}, this motion of the brane
through the bulk induces cosmology on the brane even if {\em no}
matter confined to the brane. This is sometimes called the
`mirage' effect \dc{KK} because cosmological evolution is not
necessarily sourced by the local energy density of the brane. When
matter is also included on the brane, the resulting Friedmann
equation which one obtains with this approach is identical to that
obtained when the brane is static and the bulk time dependent.
(The explicit coordinate transformation linking the two approaches
may be found in \dc{CT}.) The dark radiation term can now be
understood as being due to the motion of the brane.

Typically in both these approaches, it is assumed that the brane
divides the bulk into two identical pieces---that is, there is
$Z_2$ symmetry across the brane. This assumption can easily be
relaxed and in particular $Z_2$ symmetry will be broken if the
brane is charged and couples to a 4-form field living in the bulk
\dc{CU}. In context of the moving brane approach, one would
therefore have different cosmological constants $\Lambda_{\pm}$
and masses $M_{\pm}$ parametrising the Sch-AdS$_5$ spacetimes on
each side of the brane, and thus the brane dynamics would be
altered.  In particular, it is possible to show \dc{Kraus,Ida,CU}
that the resulting Friedman equation now has an extra dark
radiation term with energy density proportional to $[\Lambda]/a^4$
(where $[\Lambda] = \Lambda_+ -\Lambda_-$) as well as a new dark
fluid term with energy density proportional to
$[M][\Lambda]/a^{8}$ (see also section \dr{ss:linkMCJC}).

One of the questions we try to address here is the following: if
the brane is embedded in a spacetime of more than five dimensions,
what dark fluid terms are generated in the brane Friedmann
equation? To answer this question we work in the frame in which
the brane is dynamical, moving through a static or stationary
bulk. In the usual brane world scenarios, it is important to
satisfy the brane junction conditions. This reflects that the
background metric must be consistent with the presence of the
brane. Unfortunately, it is often not straightforward to apply the
junction conditions when there is more than one extra dimension
since the results typically depend on the thickness of the defect,
$\epsilon$, and are not well defined as
$\epsilon\!\rightarrow\!0$ \dc{TZ}. However, this is not fatal to
our program. For objects with codimension greater than one, it
becomes reasonable to treat them as ``test particles'' in the
background spacetime. In other words, there is no back-reaction to
solve for. This is analogous to the case of planetary orbits where
the Earth, for instance, is treated as a point particle moving in
the spacetime metric generated by the sun (see section
\dr{ss:dynSch}).

Our particular approach to brane cosmology is to consider
D3-branes in type IIB string theory. Such D3-branes are attractive
because they are stable and, by construction, matter is localised
on them. Furthermore, an action can, within certain
approximations, be derived \dc{Bachas}; it consists of the
Dirac-Born-Infeld (DBI) action plus a Wess-Zumino term. (For
slowly moving branes, the D-brane action has been used extensively
to study the properties of near extremal black holes
\dc{Malda9611125}.) The only caveat is that D3-branes are BPS
states so that one must eventually provide a prescription for
supersymmetry-breaking. As for the background in which the branes
move, this is consistently determined from the low-energy string
action. Here it is a 10D supergravity action and we consider a
Sch-AdS$_5 \times$S$_5$ bulk metric and a rotating black hole
solution, both of which can be thought of as being generated by a
stack of D3-branes. The approach we describe was coined `mirage
cosmology' (MC) and developed in depth by Kehagias and Kiritsis
\dc{KK,Kehag1,Kehag2} and extended by others \dc{Youm,NS,others}.

One of the purposes of this paper is to try to introduce MC to
cosmologists who are perhaps more familiar with 5D brane
cosmology. (As such, a part of the work presented here will follow
\dc{KK}.) The first important point is that the MC approach is
`passive'. As intimated above, the D3-brane is assumed {\em not}
to back-react on the bulk. In this sense, this approach is very
similar to that used to determine the dynamics of cosmic
topological defects. It differs from the `active' 5D case where
the junction conditions include the back-reaction of the brane on
the bulk. We dub this approach the junction conditions (JC)
approach. Secondly, notice that when there is more than one extra
dimension, the brane has much more freedom in its motion. For
example, in Sch-AdS$_5 \times$S$_5$ the brane may not only move
along the radial coordinate but also around the $S_5$. However, it
turns out that the brane angular momentum $\ell$ is conserved
around this $S_5$ (section \dr{ss:dynSch}). In section
\dr{ss:linkMCJC}, we set $\ell=0$ and discuss how the Friedmann
equation obtained from this MC approach is linked to that obtained
via the junction conditions.  In order to make this link though,
it is necessary to consider the situation in which $Z_2$ symmetry
is broken \dc{Kraus,CU} since D-branes are charged under
Ramond-Ramond fields living in the bulk \dc{Bachas}.  In section
\dr{ss:potSch} we identify a critical value of the brane angular
momentum, $\ell_c$, and discuss how the brane trajectories fall
into two very different classes depending on whether $\ell <
\ell_c$ or $\ell > \ell_c$.

A final purpose of this paper is to try to present new results on
mirage cosmology. In particular, in section \dr{s:rot}, we
consider mirage cosmology in a rotating black hole background and
comment on other work in this area. The possibility of a varying
speed of light is discussed in section \dr{ss:caus}.  In section
\dr{ss:infln} we consider brane inflation when the bulk is
generated by a `brane gas' and make other comments regarding
inflation in mirage cosmology. Finally conclusions are given in
section \dr{s:conc} where we discuss some of the limitations of
this approach to brane cosmology, perhaps most importantly, the
lack of brane self-gravity.

\section{Effective cosmology from brane motion}\dle{s:mirage}

We begin by introducing our notation and explaining briefly the
`mirage' effect.

Consider an infinitely thin $p$-brane in a $(D\!+\!1)$-
dimensional spacetime.  The following index convention will be
used to label objects:
\be
\underbrace{\underbrace{0 \overbrace{1 \ldots p}^a}_i\
\overbrace{p\!+\!1 \ldots D}^A}_{\mu}
\ee
The $D\!+\!1$ spacetime coordinates are denoted by $x^{\mu}$ with
$x^0 \equiv t$ being the time coordinate, and the background
metric $g_{\mu \nu}(x) $ has signature $(-++\ldots +)$. As the
brane moves it sweeps out a $p\!+\!1$ dimensional world-sheet
labeled by coordinates $\sigma^i$. The position of the brane in
the background spacetime is $x^{\mu}=X^{\mu}(\sigma)$ so that the
induced metric on the brane is
\be
\gamma_{ij}(\sigma) = g_{\mu \nu}(X) \frac{\partial
X^{\mu}}{\partial \sigma^i} \frac{\partial X^{\nu}}{\partial
\sigma^j}.  \dle{gamma}
\ee

We consider an infinitely long straight brane parallel to the
$x^i$-hyperplane, but free to move along the perpendicular
coordinates $x^A$. Hence a natural choice of intrinsic
coordinates\footnote{Of course any brane action must be invariant
under reparametrisations $\sigma^{i} \rightarrow
\tilde{\sigma}^{i}$, hence there is freedom to choose the
$p\!+\!1$ coordinates so as to simplify the resulting equations of
motion as much as possible.} is $\sigma^i = x^i$, and the brane
motion is described by
\be
X^{i} = x^{i}, \qquad X^{A} = X^{A}(t).
\dle{static}
\ee
This is known as the static gauge. If one wanted to study
perturbed branes, the relevant embedding would be $X^{i} = x^{i},
X^{A} = X^{A}(x^{i}) $ (see \dc{BS}).

Whether or not the induced brane metric $\gamma_{ij}$ is spatially
homogeneous and isotropic depends on the background metric. The
background line-elements considered here are either static or
stationary and take the form
\be\label{bgmet}
 ds^2 = g_{00}\; dt^2 +
\sum_{a} g_{aa} (dx^{a})^2 + 2g_{0,p+1} \; dt \; dx^{p+1} +
\sum_{A} g_{AA} (dx^{A})^2
\ee
with
\be
g_{\mu \nu} = g_{\mu \nu}(x^A).
\ee
The induced metric $\gamma_{ij}$ is then
\ba
\gamma_{00} & =&  g_{00} + 2 g_{0,p+1}
\dot{X}^{p+1} + \sum_{A} g_{AA} \dot{X}^A \dot{X}^A \nn \\
\gamma_{0a} & = & 0 \nn \\
\gamma_{ab} & = & g_{aa} \delta_{ab} \qquad \mbox{(no sum)}
\dla{gammab}
\ea
where $ \cdot = \partial/\partial t$, and the metric coefficients
are evaluated on the brane, i.e. $g_{\mu\nu}(x^A) \rightarrow
g_{\mu\nu}(X^A(t))$. It follows that $\gamma_{ij} =
\gamma_{ij}(t)$ and, in particular, that the brane is spatially
flat. To consider a curved brane we would have to assume a
different embedding (for example, see \dc{Youm}).

Now let $p=3$. We consider backgrounds for which $g_{11} = g_{22}
= g_{33} \equiv g_d$ so that the brane metric $\gamma_{ij}$ is
indeed spatially homogeneous and isotropic\footnote{If $g_{11}
\neq g_{22} \neq g_{33}$, then the brane metric is homogeneous but
anisotropic. Such a situation occurs when there is a non-zero bulk
magnetic NS field \dc{NS}. Of cosmological interest, would be
situations in which the motion of the brane (i.e.\ its expansion)
leads to isotropisation of the brane.}.  The induced line-element
on the brane is
\ba
ds^2 &=&
%\gamma_{ij}d\sigma^{i}d\sigma^{j} =
\gamma_{ij}dx^{i}dx^{j} \nn \\
& = & \gamma_{00}(t) dt^2 + g_d(X^A(t)) d{\bf x}^2 \nn \\
& \equiv & - d\tau^2 + a^2(\tau) d{\bf x}^2,
\dle{line3}
\ea
where the brane time $\tau$ is defined by
\be
d\tau = \sqrt{- \gamma_{00}(t)} dt \dle{eta}
\ee
and the brane scale factor $a(\tau)$ by
\be a^2(\tau) =
g_d(X^{A}(t(\tau))).
\dle{Sdef}
\ee
This is the `mirage' effect: the brane motion has generated a
scale factor $a(\tau)$ in the brane independently of whether or
not there is matter on the brane.  The details of $a(\tau)$ depend
both on the background, through $g_d$, and on the brane motion,
through $X^{A}(t(\tau))$. Finally the Friedmann equation is given
by
\be H^2 =
\left(  \frac{1}{a}\frac{da}{d\tau} \right)^2 = \qa
\frac{1}{|\gamma_{00}|} \frac{1}{g_d^2} \left[\sum_A
\left(\frac{\partial g_d}{\partial X^{A}} \dot{X}^{A}
\right)\right]^2 \equiv \frac{8 \pi G_4}{3} \rho_{{\rm\bf eff}}
\dle{Fried}
\ee
which defines an effective energy density.

\section{Brane action and the background metric}\dle{s:DBI}

We assume that our universe is a D3-brane in type IIB string
theory, and that the background spacetime in which it moves is
generated by all the allowed degrees of freedom. In the low-energy
limit we have a 10D supergravity action from which the bulk metric
may be determined \dc{Kelly}.  Apart from section \dr{ss:infln},
the bulk will be assumed to contain a charged stack of many
coincident D3-branes which generate, amongst other possibilities,
a Sch-AdS$_5 \times$S$_5$ bulk metric \dc{Kehag2}.

The ``universe-brane'' itself (on which, by construction, gauge
fields are confined) can also couple to many different objects and
determining its action is still an active area of research.
However, in the simplest case we can think of our universe as a
probe D3-brane whose action is given by \dc{Bachas}
\be
S = S_{DBI} + S_{WZ} = - \lambda \int d^4 \sigma \sqrt { - \det
(\gamma_{ij} + (2 \pi \alpha')F_{ij} - B_{ij} ) } - e \int \cC_4,
\dle{action}
\ee
where $\lambda$ is the brane tension, $\alpha'$ is the string
tension and $e$ is the brane charge density. The dilaton has not
been included because it is constant in the supergravity solutions
being considered. The `kinetic' term, the DBI action, is the
volume of the brane trajectory (the Nambu-Goto piece) modified by
the presence of the pull-back of the Neveu-Schwarz anti-symmetric
two-form $B_{ij}$, and worldvolume anti-symmetric
gauge fields $F_{ij}$. (The latter arise due to open strings which
may connect the probe and stack D3-branes.)  Thus, for example, if
there is radiation on the brane, $F_{ij} \neq 0$, and the brane
dynamics will be altered relative to the case of a brane with no
radiation.\footnote{This is exactly the same effect as in the case
of current carrying cosmic strings. With no current, the string
action is the NG action.  With a current, the action is changed
and it may lead to very different cosmic string dynamics---for
example, stable loops called vortons may now be formed \dc{US}.}
The modified dynamics will in turn change the Friedmann equation
(as explained in section \dr{s:mirage}) which will thus contain
terms reflecting the presence of the radiation \dc{KK,NS} (see
also section \dr{ss:radSch}).

Note, however, that the brane action as it stands does not allow
for arbitrary matter content. Thus, as presented so far, MC cannot
provide a full account of the evolution of our universe. (It
would, after all, be unbelievable if our current cosmology, based
upon 4D gravity and the local matter density, could be emulated
solely by the motion of a brane in a higher-dimensional
background.) However, MC may well have a role to play where our
understanding is not well established, namely at early and late
(future) times. Additionally, (\dr{action}) provides a springboard
for more phenomenological approaches. We return to these themes in
section \dr{ss:infln}.

The Wess-Zumino term in (\dr{action}) is required since the probe
D3-brane is charged under Ramond-Ramond gauge fields living in the
bulk.  Here it takes the simple form given in (\dr{action})
because we assume that the stack is the sole source of RR fields.
Thus the only contribution is from a 4-form $\cC_4$ and
\be
S_{WZ} = - e \int \cC_4 = -e \int \frac{1}{4!} C_{\mu \nu \rho
\tau} \frac{\partial X^{\mu}}{\partial\sigma^{i}} \frac{\partial
X^{\nu}}{\partial\sigma^{j}}\frac{\partial
X^{\rho}}{\partial\sigma^{k}}\frac{\partial
X^{\tau}}{\partial\sigma^{\ell}} d\sigma^{i} d\sigma^{j}
d\sigma^{k}d\sigma^{\ell},
\dle{WZ}
\ee
where the gauge field $C_{\mu \nu \rho \tau}$ is obtained from the
corresponding field strength $F_{\alpha \mu \nu \rho \tau}$ (which
off the brane and for $r\!>\!0$ is a solution of $ \nabla^{\alpha}
F_{\alpha \mu \nu \rho \tau} = 0$).
%\dc{Kehag2}).
By virtue of the coordinate choice (\dr{static}), we have assumed
that the probe brane is parallel or anti-parallel to the stack.
Supersymmetry of the total system remains unbroken only in the
parallel case when the brane is BPS implying that $e = \lambda$
\dc{Bachas}. The anti-parallel case corresponds to the probe being
an anti-brane and to $e=-\lambda$.  However, in order to make
comparisons with more phenomenological brane world scenarios and
ones using the JCs, we will write more generally
\be
e = q \lambda.
\dle{chargedef}
\ee

\section{Mirage cosmology in Schwarzschild-AdS$_5 \times
S_5$}\dle{s:Sch}

A particularly illustrative background in which to apply the
mirage cosmology approach is Sch-AdS$_5 \times$S$_5$, since if one
dropped the $S_5$ piece it would correspond to the background
metric used in the moving brane approach to 5D brane cosmology described
in the introduction \dc{Kraus}.  Thus the dynamics of the brane
around the $S_5$ should give an indication of which dark fluid
terms are generated in models with more than one extra
dimension\footnote{Notice that the induced metric on the brane
corresponds
to a flat universe and that the brane does not wrap around the
$S_5$ by virtue of the embedding (\dr{genmet}). One could choose instead
to wrap the
brane around part of the $S_5$ so leading to a closed universe
\dc{Youm}.}. The
Sch-AdS$_5 \times$S$_5$ metric is given by
\be
ds^2 = \frac{r^2}{L^2} \left( -f(r) dt^2 + dx_1^2 + dx_2^2 +
dx_3^2 \right) + \frac{L^2 dr^2}{f(r) r^2 } + L^2 d\Omega_5^2
\dle{AdS}
\ee
where
\be
f(r) = 1- \left( \frac{r_0}{r} \right)^4
\ee
and $d\Omega_5^2 = h_{IJ}(\phi) \phi^I \phi^J$ is line element of
the unit 5-sphere described by coordinates $\phi^I$, $I=0\ldots 5$. The
metric satisfies the 10D Einstein equations with $\Lambda \equiv -16/L^2$
and $r_0^4\equiv2 G M L^2$ gives
the
black hole mass $M$, with $G$ the
10D Newton constant. In this background the radial position of
the brane determines the brane scale factor $a$ since from
(\dr{Sdef})
\be
a = \frac{r}{L}~.
\dle{adef}
\ee
We take $r_0/L\leq a \leq \infty$, though it would be interesting to
determine what happens for a brane that crosses the black hole horizon.

In order to obtain $a(\tau)$ the brane dynamics $r(\tau)$ must be
calculated.  Initially (section \dr{ss:dynSch}) we assume that
there is no radiation on the brane, $F_{ij}=0$, and turn off the
bulk NS fields, $B_{ij} = 0$. Then the Friedmann equation which
follows from (\dr{adef}) will contain only dark fluid terms. Some
of these terms (sections \dr{ss:dynSch} and \dr{ss:linkMCJC}) will
be seen to be the familiar dark fluid terms of 5D brane worlds
mentioned in the introduction. However, other terms arise from the
non-trivial dynamics of the brane around the $S_5$ and they can
lead to some interesting effects (section \dr{ss:potSch}). In
section \dr{ss:linkMCJC} we define precisely the link between this
MC approach to brane world cosmology and the junction condition
approach used in 5D. Finally in section \dr{ss:radSch}  we
consider briefly the case of non-zero $F_{ij}$. Parts of sections
\dr{ss:dynSch} and \dr{ss:radSch} follow closely reference
\dc{KK}.

\subsection{Brane dynamics with no matter}\dle{ss:dynSch}

To maintain some generality we write the bulk metric line element as
\be
ds^2 = g_{00}(r) dt^2 + g_d(r) d{\bf x}^2 + g_{rr}(r) dr^2 + g_s(r)
d\Omega_5^2
\dle{genmet}
\ee
which includes the Sch-AdS$_5 \times$S$_5$ metric of (\dr{AdS}).
The only non-zero component\footnote{Actually, the self-duality
condition for the field strength for $p=3$ means the 4-form will
also have non-zero components in the $S_5$-directions, but these
do not contribute to the WZ term in the static gauge.} of $C_{\mu
\nu \rho \tau}$ is then $C_{0123}(r) \equiv C_4(r)$. Thus in the
gauge (\dr{static}), $\cC_4 = C_4(r) d^4x$ and, with $F_{ij} =
B_{ij} = 0$, the action (\dr{action}) defines a dimensionless
Lagrangian ${{\cal L}}$ through
\be
S = - \lambda \int d^4 x \sqrt { - \det ({\gamma_{ij}} ) }
- q \lambda \int d^4 x \; {C}_4
 \equiv   \lambda V_3 \int dt {{\cal L}},
\dle{SDBIdef}
\ee
where $V_3 = \int d^3x$. Using (\dr{gamma}) gives
\be
\cL = - \sqrt{-g_d^{3}\gamma_{00}} - q  {C}_4 \equiv   - \sqrt{\cA
+ \cB \dot{r}^2 + \cC h_{IJ}\dot{\phi}^I \dot{\phi}^J} + \cE
\dle{Lagdef}
\ee
with
\be
\cA =-g_d^3 g_{00}, \qquad \cB = -g_d^3 g_{rr},  \qquad \cC =
-g_d^3 g_s, \qquad \cE =  -q C_4.
\dle{abcedef}
\ee
By inspection, since ${\cal L}$ is not explicitly time dependent
and the $\phi$-dependence is confined to the kinetic term for
$\dot{\phi}$, the brane geodesics are parametrised by a conserved
energy $E$ and an angular momentum $\ell^2$ given, respectively, by
\be\dle{firstints}
E = \frac{\partial {\cal L}}{\partial \dot{r}} \dot{r} +
\frac{\partial {\cal L}}{\partial \dot{\phi}^I} \dot{\phi}^I
- {\cal L}, \qquad \qquad \ell^2  =  h^{IJ}\frac{\partial
{\cal L}}{\partial \dot{\phi}^I} \frac{\partial {\cal L}}{\partial
\dot{\phi}^J}.
\ee

Solving these expressions for $\dot{\phi}$ and $\dot{r}$ gives
\be
h_{IJ} \dot{\phi}^I \dot{\phi}^J  =   \frac{\cA^2
\ell^2}{\cC^2(E+\cE)^2}~, \qquad \dot{r}^2  =  -
\frac{\cA}{\cB} \left[ 1 + \frac{\cA}{\cC} \frac{ (\ell^2 -
\cC)}{(E+\cE)^2} \right].
\dle{dotr}
\ee
The brane time $\tau$ is then obtained on substitution of
(\dr{dotr}) into (\dr{eta}):
\be
d\tau^2 = \frac{1}{g_d^3}(\cA + \cB \dot{r}^2 + \cC
h_{IJ}\dot{\phi}^{I} \dot{\phi}^{J}) dt^2  =
\frac{\cA^2}{(E+\cE)^2} \frac{1}{g_d^3} dt^2
\dle{teta}
\ee
and the Friedmann equation (\dr{Fried}) becomes
\be
H^2 = - \frac{g_d (g_d')^2}{4 \cA \cB \cC} \left[ \cA (\ell^2 - \cC) +
\cC (E+\cE)^2
\right].
\dle{H2}
\ee

The specific forms of $\cA,\cB,\cC,\cE$ and $a$ for the
Sch-AdS$_5 \times$S$_5$ metric (\dr{AdS}) are
\ba
\cA = \frac{r^8}{L^8} f  = a^8 \left( 1-\frac{X^4}{a^4} \right), &
\qquad & \cB = -  \frac{r^4}{L^4 f} = - {a^4 } \left(
1-\frac{X^4}{a^4} \right)^{-1},
\nn
\\
\cC = - \frac{r^6}{L^4} = - a^6 L^2 , & \qquad & \cE = q \left(
\left(\frac{r}{L}\right)^4 - \frac{X^4}{2} \right)= q \left( a^4 -
\frac{X^4}{2} \right)
\dla{defSAdS}
\ea
where $X = r_0/L$ and we have used the expression for the 4-form
in Sch-AdS$_5 \times$S$_5$ \dc{KK}
\be
{C}_4(r) = - \frac{r^4}{L^4} + \frac{r_0^4}{2L^4}
\dle{4form}.
\ee
Finally, the Friedmann (\dr{H2}) equation is
\be
H^2 =
 \frac{q^2-1}{L^2} + \frac{X^4 L^2}{a^4} +
 L^6   \left( \frac{\tilde{E}}{a^4} \right)
\left( \frac{\tilde{E}}{a^4} + \frac{2q}{L^4} \right) +
\, \ell^2\, \frac{ L^4}{a^6} \left(  \frac{X^4 L^2}{a^4} -
\frac{1}{L^2} \right),
\dle{res}
\ee
where we now rescaled the scale factor $a$ by a factor of
$L$ to give it dimensions of length and the constant
part of $\cE$, essentially electrostatic energy, has been absorbed into
the energy so that $\tilde{E}
= E - qX^4/2$.  Note that the first term is the effective cosmological
constant on the brane but that this vanishes when $q=\pm
1$.

The dependence of this Friedmann equation on $\ell$ will be
discussed in subsection \dr{ss:potSch} where we will comment on
the final term of (\dr{res}) which contributes a negative energy
density for $r>r_0$. We now focus on the case $\ell = 0$.

\subsection{$\ell = 0$: `mirage' versus brane world cosmology}\dle{ss:linkMCJC}

When $\ell=0$ the D3-brane has no dynamics about the $S_5$
and hence its motion is effectively
constrained
to a Sch-AdS$_5$ bulk with metric
\be
ds_5^2 =  \frac{r^2}{L^2} \left( -f(r) dt^2 + dx_1^2 + dx_2^2 +
dx_3^2 \right) + \frac{L^2 dr^2}{f(r) r^2 }.
\dle{AdS5}
\ee
It is straightforward to work out the Friedmann equation resulting
from the MC approach in this case: it is given in (\dr{res}) where
one must set $\ell=0$. (Note that the cosmological constant
corresponding to (\dr{AdS5}) is now $\Lambda = -6/L^2$ and that
$r_0^4\equiv2 G_5 M L^2$ where $G_5$ is the 5D Newton constant. In
obtaining the Friedmann equation we use (\dr{4form}) which also
holds in 5D \dc{BS}.) However, as was discussed in the introduction, the
brane dynamics---including the back-reaction of the brane on the
bulk---are also known in this case (i.e.\ with bulk metric
(\dr{AdS5})) from the JC approach \dc{Kraus,Ida,CU}. The purpose
of this section is to compare these two Friedmann equations.  Are
they related in any way? And if so, how do the different
parameters relate to one another? In other words, when there is
only one extra dimension, how does the `test brane' MC approach
compare with the `exact' JC approach?

Before making this comparison, note three important points.
Firstly, since the D3-branes we are considering in this paper are
charged and couple (minimally) to a 4-form field living in the
bulk, one must consider, in the JC approach, a setup in which
$Z_2$ symmetry is broken---see comments in the introduction and
\dc{CU}.  Thus the brane, with charge $e_4$ say, divides two
different regions of Sch-AdS$_5$ with cosmological constants and
masses $\Lambda_\pm,M_\pm$.  Secondly, in deriving (\dr{res}) we
have set $F_{ij} = B_{ij}=0$ and so there is no matter on the
brane, thus we must set $\rho=p=0$ in the JC approach.  Finally,
we have considered a flat brane, so $k=0$ in the JC approach.

Once these conditions are imposed, the resulting Friedmann
equation calculated in \dc{CU} using the junction conditions
depends on $G_5$ and $e_4$ as well as five other dimensionful
quantities: the brane tension $\lambda$---which is the same as
that in (\dr{action})---and $\Lambda_\pm, M_\pm$. In fact the
combinations which appear are $\langle M \rangle$, $[M]$, $\langle
\Lambda \rangle$ and $[\Lambda]$ where $\langle x
\rangle\equiv (x_+\!+\!x_-)/2$
and $[x]\equiv x_+\!-\!x_-$.
A final important identity relates the force on the brane $e_4
\langle F \rangle$ to the jump in cosmological constant \dc{CU}:
\be
[\Lambda] = 6 \pi^2 e_4 G_5 \langle F \rangle.
\dle{CUparam}
\ee
(Here $F$ is defined through the physical 5-form field strength
$F_{\mu \nu \rho \sigma \tau} = F \epsilon_{\mu \nu \rho \sigma
\tau}$ corresponding to the bulk gauge field $A_{\mu \nu \rho
\sigma}$. From the 5D SUGRA equations of motion it is
straightforward to show that $F = K_1 r^3/L^4$, where the constant
$K_1$ is dimensionless, and that $A_{0123}$ is, up to a
multiplicative constant, just $C_{0123}$ of (\dr{4form}) used in
the MC approach.  Furthermore, from the equations
of motion, the effective cosmological constants are easily seen to
be $\Lambda_\pm = \Lambda \pm K_2 F_{\pm}^2$ where $K_2$ is
another numerical constant and $\Lambda = -6/L^2$.)  Notice from
(\dr{CUparam}) that if the brane is uncharged, $e_4=0$, then
$[\Lambda]=0$ so that there is no force on the brane.  Finally the
resulting Friedmann equation is \dc{CU}
\be
H^2 = \frac{\Lambda_4}{3} + \frac{2G_5\langle M \rangle}{a^4} +
\left( \frac{3}{8 \pi \lambda} \right)^2 \frac{[M]}{a^4} \left(
\frac{[M]}{a^4} + \pi^2 e_{4} \langle F \rangle \right),
\dle{CU}
\ee
where $a$ is the dimensionful scale factor, and the effective
cosmological constant $\Lambda_4$ is given by
%\footnote{The brane tension $\lambda$ is related to the parameter
%$T_{\infty}$ of \dc{CU} by $T_{\infty} = 4 \lambda/3\pi$.}
\be
\frac{\Lambda_4}{3} = \frac{\langle \Lambda \rangle}{6} + \qa
\left( \frac{8 \pi \lambda}{3} \right)^2 G_5^2 + \qa \left(
\frac{3}{8 \pi \lambda} \right)^2 \left( \pi^2 e_{4} \langle F
\rangle \right)^2.
\dle{lambdaCU}
\ee
If $Z_2$ symmetry is imposed, i.e.\ $[M] = [\Lambda] =0$, then
(apart from the cosmological constant term) the Friedmann equation
(\dr{CU}) contains only the familiar dark radiation term coming
from the electric part of the Weyl tensor. With no $Z_2$ symmetry
there is the extra dark radiation term plus a contribution $\sim
a^{-8}$, as mentioned in the introduction.

How does this Friedmann equation (\dr{CU}) compare with the
Friedmann equation (\dr{res}) obtained from the DBI action when
$\ell=0$?  That equation is
\be
H^2=\frac{q^2-1}{L^2} + \frac{2G_5 M}{a^4} +
 L^6  \left( \frac{\tilde{E}}{a^4} \right)
\left( \frac{\tilde{E}}{a^4} + \frac{2q}{L^4} \right).
\dle{mirage}
\ee

Notice first that (\dr{mirage}) and (\dr{CU}) have a very similar
form, and in particular exactly the same scale factor dependence.
The familiar dark radiation term of $Z_2$ symmetric brane worlds
is also found in the MC approach---it is the term $2G_5
M/a^4$---and the two approaches are seen to lead to the same
`dark' fluids on the brane.

Next one can compare the coefficients of the various terms in
equations (\dr{mirage}) and (\dr{CU}). How are the four parameters
$(q,\tilde{E},M,\Lambda)$ parametrising the geodesic motion of the
test brane in Sch-AdS$_5$ related to the five parameters
$(e_4,M_\pm,\Lambda_{\pm})$ in the JC approach? Clearly this
identification will force two of these last six parameters to be
related. However, before making this identification note one final
important point: it is the application of junction conditions
which gives rise to the term proportional to $G_5^2$ in
(\dr{lambdaCU}), and we should not expect such a term in the MC
approach. Comparing (\dr{CU}), (\dr{lambdaCU}) and (\dr{mirage})
this is indeed verified.  Furthermore, since both $q$ and $e_4$
are brane charges, we expect $q \propto e_4$ so that one deduces
that
\be
\Lambda = \langle \Lambda \rangle  ,  \qquad \qquad M =  \langle M
\rangle.
\dle{resa}
\ee
Thus, for example, the mass $M$ of the Schwarzschild black hole
appearing in the MC approach must be identified with the average
mass $ \langle M \rangle$ in the JC approach. Then, since $L$ is
independent of $M$, it ought to be independent of $[M]$, and this
forces $L^4\propto \lambda^{-1}$. Here the constant of
proportionality is arbitrary because of the freedom in how $E$ is
defined and because $q$ is also scaled by the multiplicative
constant relating $A_{\mu\nu\rho\sigma}$ and
$C_{\mu\nu\rho\sigma}$. Thus we are free to write
\be
\frac{\tilde{E}}{L} = [M]  \qquad  \frac{2q}{L^5} = \pi^2 e_4
\langle F \rangle
\dle{resb}
\ee
which forces
\be
\langle \Lambda \rangle = -4 \sqrt{6 \pi \lambda}.
\dle{resss}
\ee

\subsection{Effects of angular momentum}\dle{ss:potSch}

We now return to the full 10D case of section \dr{ss:dynSch} and
consider non-zero angular momentum, $\ell \neq 0$.  Now the
Friedmann equation (\dr{res}) has two extra contributions. The
first is proportional to $a^{-6}$ and is characteristic of an
equation of state $w\!\equiv\!p/\rho\!=\!1$. However it
contributes with a negative energy density in (\dr{res}). The
second is proportional to $a^{-10}$ which would correspond to
matter with equation of state $w\!=\!7/3$.  To understand the
effect of these terms, it is helpful to construct an effective
potential for the brane motion as a function of the radial
coordinate $r$.

Our approach is the same as that used when considering planetary orbits:
the constants of the motion are used to eliminate all but the radial
degree of freedom. Then the 1D equation of motion follows from the
Lagrangian,
$L=\ha \dot{r}^2 - V_{\bf eff}(r)$, where $V_{\bf eff} \equiv E-\ha
\dot{r}^2$.

Here, two effective potentials can be constructed.  The
first, $V^t_{\bf eff}$, determines the brane dynamics as seen by
an observer outside the brane whose time coordinate is $t$:
\ba
V^t_{\bf eff}(r,\ell,E) &\equiv & E - \ha \dot{r}^2\; \;  =
\; \; E + \frac{\cA}{2\cB} \left[ 1 + \frac{\cA}{\cC} \frac{
(\ell^2 - \cC)}{(E+\cE)^2} \right]
\nn
\\
 &=& E - \ha\left(\frac{r}{L}\right)^{4} f^2 \left[
1 - r^2 f \frac{(\ell^2 L^4 + r^6)}{(\tilde{E}L^4 + q r^4)^2}
\right],
\dla{Vefft}
\ea
where we have used (\dr{dotr}) followed by (\dr{defSAdS}). In a
similar way, the second potential $V^{\tau}_{\bf eff}$, which is
defined for an observer living on the brane, is given by
\ba
V^{\tau}_{\bf eff}(r,\ell,E) &\equiv&  E - \ha\left(
\frac{dr}{d\tau} \right)^2 = E +  \frac{g_d^3}{2\cA \cB \cC}
\left[ \cC(E+\cE)^2 + \cA (\ell^2 - \cC) \right]
\nn
\\
& =& E + \ha\left(\frac{L}{r}\right)^{6} \left[ f
\frac{r^2}{L^4} \left( \ell^2 + \frac{r^6}{L^4} \right) - \left(
\tilde{E} + q \frac{r^4}{L^4} \right)^2 \right].
\dla{Veffeta}
\ea
This second potential is more relevant for cosmology. Initially, however,
we study both potentials, focusing on BPS branes for
which $q=1$ (for $q \neq 1$ see \dc{BS}).

First consider some properties of $ V^t_{\bf eff}$.
Since $f=0$ at the horizon ($r=r_0$), it follows from (\dr{Vefft})
that $V^t_{\bf eff}(r_0) = E$ and $\partial V^t_{\bf eff} /
\partial r \left. \right|_{r=r_0} = 0$.  Hence the potential has a
turning point at the horizon.  Also $V^t_{\bf eff}(r\!\rightarrow\!
\infty) = 0$.  Thus only when $E=0$ does the
brane have zero kinetic energy at infinity. The behaviour of
$V^t_{\bf eff}$ between the horizon and infinity depends on the
size of $\ell^2$. This is illustrated in figure \dr{fig:Vefft} where we
have introduced the rescaled quantities: $\hat{V}_{\bf
eff}^t\!=\!V_{\bf eff}^t L^4/r_0^4$, $\hat{E}\!=\!E L^4/r_0^4$,
$\hat{r}\!=\!r/r_0$, $\hat{L}\!=\!L/r_0$ and $\hat{\ell}\!=\!\ell
L^2/r_0^3$.
\begin{figure}[ht]
\begin{center}
\psfrag{r}[b]{\hspace{8cm}$\hat{r}$}
\psfrag{1.5}[]{\hspace{0.3cm}$\hat{V}_{\bf eff}^t$}
\psfrag{2.5}[]{}
\psfrag{0.5}[]{}
\psfrag{7}[]{}
\includegraphics[]{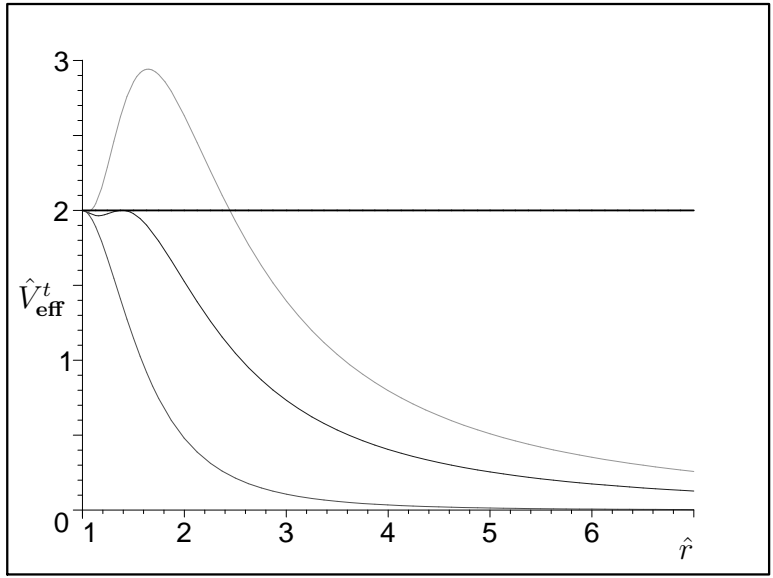}
\caption{The rescaled
effective potential $\hat{V}^t_{\bf eff}$ for $q\!=\!1$. The parameters
are $\hat{E}\!=\!2$ and $\hat{L}\!=\!1$. The lower curve has
$\hat{\ell}\!=\!0$, the
upper one $\hat{\ell}\!=\!5$ and the middle one
$\hat{\ell}\!=\!\hat{\ell}_{c}\!=\!3.49$.} \label{fig:Vefft}
\end{center}
\end{figure}

\begin{figure}[ht]
\begin{center}
\psfrag{1.96}[]{}
\psfrag{2.04}[]{}
\psfrag{2.08}[]{\hspace{0.8cm}$\hat{V}_{\bf eff}^t$}
\psfrag{1.6}[]{}
\psfrag{r}[b]{\hspace{8cm}$\hat{r}$}
\includegraphics[]{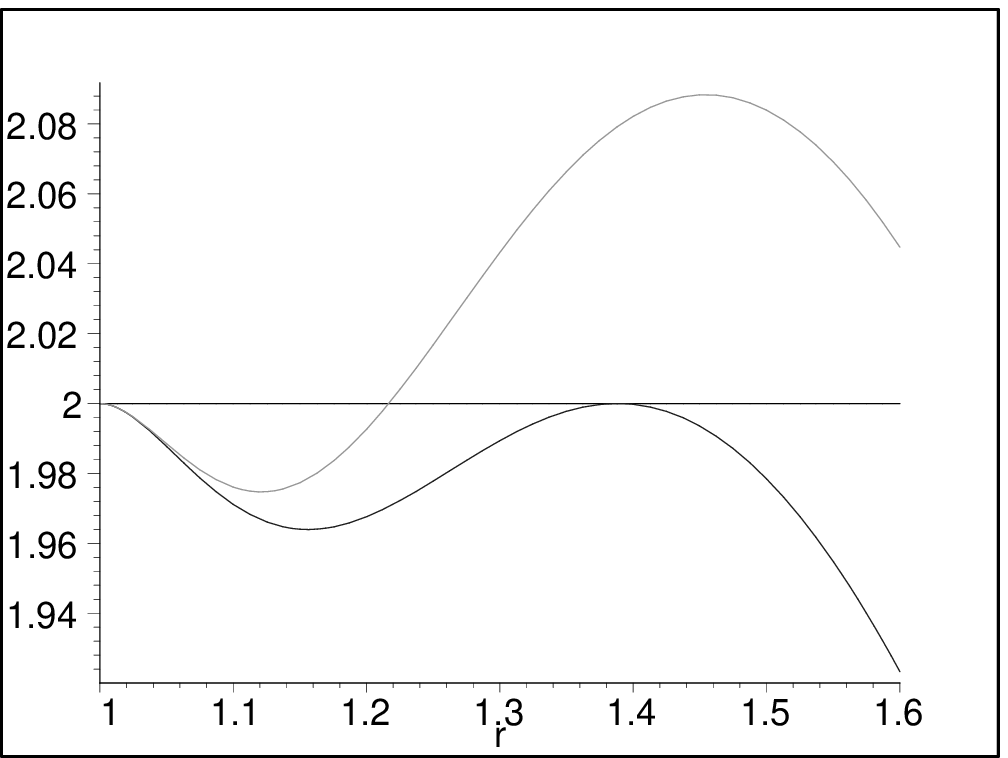}
\caption{Detail of
the rescaled effective potential $\hat{V}^t_{\bf eff}$. The parameters
$\hat{E}$
and $\hat{L}$ take the same values as in figure \dr{fig:Vefft}. The lower
curve has
$\hat{\ell}\!=\!\hat{\ell}_{c}\!=\!3.49$ and the upper one
$\hat{\ell}\!=\!3.7$. The critical radius is $\hat{r}_c\!=\!1.39$.}
\label{fig:Vefftdetail}
\end{center}
\end{figure}
If $\ell = 0$ (the lower line in the figure) and the brane is
moving radially inwards, it reaches the horizon as $t \rightarrow
\infty$ where it is `absorbed' by the black hole.  Alternatively,
a brane initially moving radially outwards escapes to infinity.

If the brane has a large angular momentum $\ell$, as in the upper
curve of figure \dr{fig:Vefft}, then a centrifugal potential barrier
forms. Thus if the brane initially moves
inwards from infinity, it bounces back at a given radius to move
back out to infinity. On the other hand, the brane could also be
trapped in the small region near the horizon (see figure
\dr{fig:Vefftdetail}). Suppose that a brane moves radially
outwards in this region: it continues moving outwards until it is
reflected off the potential barrier eventually being
absorbed by the black hole.

There is a critical value of the angular momentum $\ell_{c}$ and
corresponding critical radius $r_{c}\!>\!r_0$ for which $V^t_{\bf
eff}(r_c) = E$ and $\partial V^t_{\bf eff} / \partial r \left.
\right|_{r=r_c} = 0$. (See the middle curve in figure \dr{fig:Vefft} and
the lower one in figure \dr{fig:Vefftdetail}). With this angular momentum
the brane may reach a stable circular orbit with radius $r_c$. The
expression for $\ell_{c}$ is given in the appendix.

Now consider the effective potential $V^{\tau}_{\bf eff}$
of equation (\dr{Veffeta}).  This describes the brane trajectory
as a function of brane-time $\tau$, and since $r(\tau) = a(\tau)$,
also the behaviour of the scale factor. Notice that $V^{\tau}_{\bf
eff}(r_0) = E-\ha X^{-6} (E + \ha X^4)^2 \neq 0$ is
$\ell$ independent, and that
% In other words, $\spr{a}$ does not vanish at the
%horizon.
as $r\rightarrow \infty$, $V^{\tau}_{\bf eff} \rightarrow E$.
Hence in this limit $(dr/d\tau)^2=0$, which reflects the fact that there
is no cosmological constant in this case
(of $q=1$). Furthermore, observe that the coefficient of the
$\ell^2$-term is positive and is given by $f L^2 /2 r^4$; this is
responsible for the centrifugal barrier.

The
behaviour of $V^{\tau}_{\bf eff}$ as a function of $r$ is
shown in figure \dr{fig:Veffeta}.
\begin{figure}[ht]
\begin{center}
\psfrag{r}[b]{\hspace{9.3cm}$\hat{r}$}
\psfrag{1}[]{\hspace{-0.6cm}$\hat{V}_{\bf eff}^{\tau}$}
\includegraphics[]{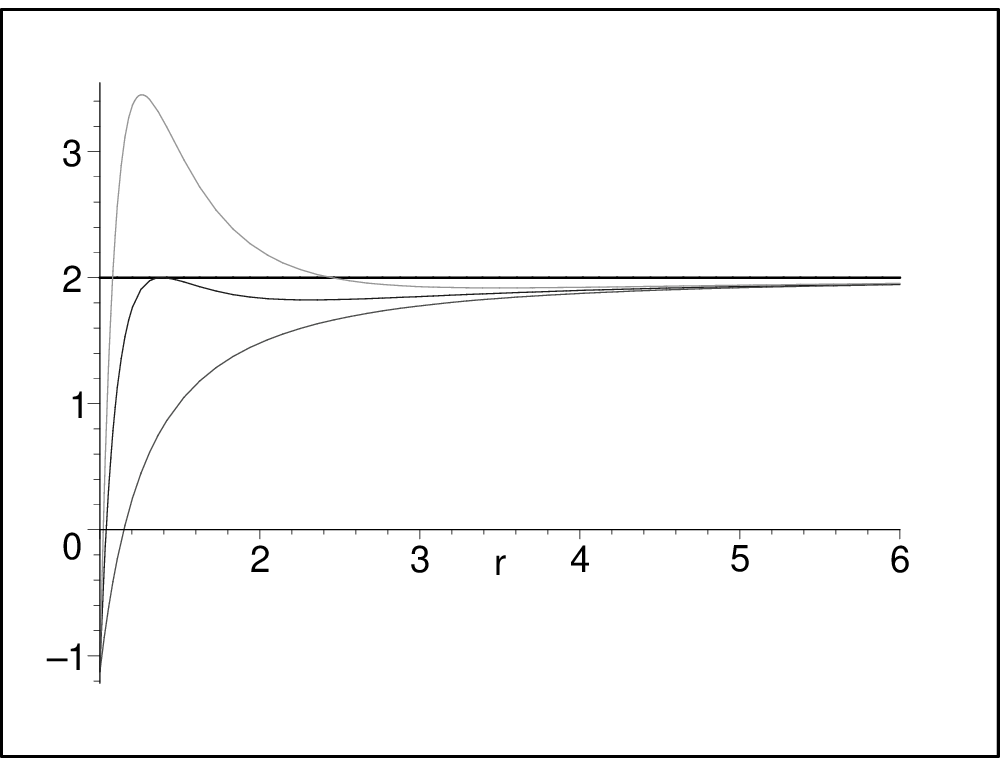}
\caption{The rescaled
effective potential $\hat{V}^{\tau}_{\bf eff}$ for the same values of
the parameters as in figure 1. } \label{fig:Veffeta}
\end{center}
\end{figure}
Again one identifies three regimes:
\begin{itemize}
\item  $\ell < \ell_c$.  The universe either expands or
contracts forever.  Expansion/contraction depends on whether the
brane initially moves radially outwards/inwards.
\item $\ell>\ell_c$.  Here there are two possibilities.  {\it i)}
The universe initially contracts---corresponding to the brane
moving in from infinity---before bouncing off the centrifugal
barrier and starting a period of expansion. {\it ii)} The brane
moves radially outwards from the horizon, expanding at the same
time, and then bounces off the centrifugal barrier. It then
contracts before it terminating its life after some finite brane
time inside the black hole---a `black crunch'.
\item $\ell = \ell_c$.  If the brane moves radially inwards from
infinity, the universe will contract but the rate of contraction
will decrease until, after an infinite amount of time, the scale
factor takes the constant value $a= r_c$.
If, on the other hand, the brane moves radially outwards from the
horizon then it expands but again the scale factor reaches the
value $a=r_c$.
\end{itemize}

Allied with the question of brane dynamics is the question of
brane initial conditions. Clearly for an expanding solution we
require the brane to moving outwards from the black hole, but this
begs the question of how the brane came to be in this state. One
interesting idea is to suppose the brane is Hawking emitted from
the black hole. However, the probability of such an event is
thought to be extremely low \dc{Malda9611125}.

\subsection{Radiation on the brane}\dle{ss:radSch}

So far we have focused on the dynamics of branes with no matter on
them. In this subsection we comment very briefly on branes with
radiation, that is $F_{ij} \neq 0$ (see \dc{KK}).  The important
point is that the Friedmann equation resulting from the MC
approach contains no $\rho_{rad}^2$ terms, but only terms linear
in $\rho_{rad}$ (the energy density in radiation on the brane).
This contrasts with the JC approach where the back-reaction of the
brane on the bulk metric is responsible for the $\rho_{rad}^2$
terms \dc{Binetruy,Kraus}.

As mentioned in section \dr{ss:dynSch}, if electromagnetic fields
are present on the brane their non-zero energy density affects the
brane dynamics via the action (\dr{action}).  Following \dc{KK} we
now determine the effect of a uniform electric field $\langle {\bf
E}^2 \rangle$ on the brane dynamics and hence its contribution to
the Friedmann equation.

From (\dr{action}), the Lagrangian is now
\be
{{\cal L}} = - \sqrt{\cA + \cB \dot{r}^2 + \cC h_{IJ}\dot{\phi}^I
\dot{\phi}^J - {\bf E}^2 g_d^2} + {\cE} \equiv - \sqrt{Z} + \cE
\dle{LdefE}
\ee
where ${\bf E}^2 = 2 \pi \alpha'E_i E^{i}$ and $E_0 = -\dot{A}_i$ in
the gauge $A_0=0$.  The equation of motion
for $E_i$ gives \dc{KK}
\be
{\bf E}^2 = \frac{\mu^2 Z}{g_d^4},
\ee
where $\mu^2 =2 \pi \alpha'\mu_i \mu^{i}$ and $\mu_i$ are
integration constants. Solving for ${\bf E}^2$ yields
$$
{\bf E}^2g_d^2 = \left( \frac{\mu^2}{\mu^2 + g_d^2} \right)\left(
\cA + \cB \dot{r}^2 + \cC h_{ij}\dot{\phi}^I \dot{\phi}^J \right),
$$
so that (\dr{LdefE}) becomes
\be
{{\cal L}} = - \sqrt{\cA' + \cB'\dot{r}^2 + \cC' h_{IJ}\dot{\phi}^I
\dot{\phi}^J } + {\cE},
\dle{LdefEbis}
\ee
where $\cA'= \cA \left({ 1 + \mu^2 g_d^{-2} } \right)^{-1}$ and
identical relations hold for $\cB'$ and $\cC'$.  Hence the
expressions for $\dot{r}^2$ and $h_{IJ}\dot{\phi}^I \dot{\phi}^J$
are just as in (\dr{dotr}), but with $\cA \rightarrow \cA'$, etc.
Furthermore, it is straightforward to show that $dt^2 = g_d^3
\cA{}^{-2} (E+\cE)^2 (1 + \mu^2 g_d^{-2}) d\tau^2$ so that the
Friedmann equation including the effects of radiation is
\be
H_{{\bf total}}^2 = H^2 + H_{\langle {\bf E}^2\rangle}^2,
\ee
where $H^2$, which is independent of $\mu^2$, is given in equation
(\dr{teta}), and
\ba
H_{\langle {\bf E}^2\rangle}^2 &=& - \frac{1}{4 \cA \cB \cC}
\frac{(g_d')^2}{g_d} \mu^2 \left[ \cC (E+\cE)^2 + \cA \ell^2
\right]
\nn
\\
&=& \rho_{rad} \left( qL + \frac{\tilde{E} L^5}{a^4} \right)^2 +
\ell^2 \rho_{rad} \left( \frac{L^{8}}{a^{6}} \right) \left(
\frac{X^4 L^2}{a^4} - \frac{1}{L^2} \right).
\ea
Here, since $\rho_{rad} \equiv \mu^2/a^4$ is energy density in
radiation,
%, as in (\dr{res}), $a(\tau)$ is the scale factor.  Since the
it would appear that the 4D Newton constant should be identified
with
\be
\frac{8 \pi G_4}{3} = q^2 L^2 = \frac{16 q^2}{(-\Lambda)}.
\ee

Hence, to summarise, the final result (now setting $q=1$) is that
\ba
H^2 &=& \frac{8 \pi G_4}{3} \rho_{rad} + \Delta
\nn
\\
\Delta & = & \frac{X^4 L^2}{a^4} + (\rho_{rad} L^4 + 1) \left[L^6
\left( \frac{\tilde{E}}{a^4} \right) \left( \frac{\tilde{E}}{a^4} +
\frac{2}{L^4}  \right) + \ell^2 \frac{L^4}{a^6} \left(\frac{X^4
L^2}{a^4} -\frac{1}{L^2} \right) \right]
\dla{Deltadef}
\ea
As commented above, this is linear in $\rho_{rad}$.  If equation
(\dr{Deltadef}) were to describe a realistic cosmology then one
could now proceed to try to constrain $\ell, \tilde{E}, L$ and
$r_0$ by nucleosynthesis constraints.

\section{Mirage cosmology in a rotating black hole bulk}\dle{s:rot}

An aspect of the Sch-AdS$_5 \times$S$_5$ background
which simplifies considerations is that the scale factor is
just the
radial distance of the brane from the black hole.
When the brane moves in other backgrounds the expression for $a(\tau)$ is
generally more complicated: recall from
(\dr{Sdef}) that $a^2(\tau) = g_d(X^A(\tau))$.

As an example of this, we consider MC in a rotating black hole
background. This supergravity solution was constructed in
\dc{KLT}, and brane dynamics and thermodynamics in this
background were studied in \dc{Cai,Cai2}. The possibility of a
varying speed of light effect in MC was addressed in
\dc{Kehag2,Alex}. In this section we introduce a general formalism
for studying MC in the background of a rotating source.  We
clarify and correct aspects in recent literature. Additionally, we
consider the dark fluid terms that arise in this model. We comment
more fully on varying speed of light effects in section
\dr{ss:caus}.

\subsection{Background metric}\dle{ss:metricrot}

The metric for the rotating black hole solution is \dc{KLT}
\ba
ds^2 & = & \frac{1}{\sqrt{f}} \left( -h dt^2 + dx_1^2 + dx_2^2 +
dx_3^2 \right) + \sqrt{f} \Big [ \frac{dr^2}{\tilde{h}} -
\frac{4ml\cosh \alpha}{r^4 \Delta f} \sin^2 \theta dt d\phi
\nn
\\
& + &  r^2 (\Delta d\theta^2 + \tilde{\Delta} \sin^2 \theta
d\phi^2 + \cos^2 \theta d\Omega_3^2) \Big ]
\dla{rotD}
\ea
where
\ba
f &=& 1 + \frac{2m\sinh^2 \alpha}{r^4 \Delta} \equiv  1 +
\frac{\tilde{R}^4}{r^4 \Delta}
\nn
\\
\Delta &=& 1 + \frac{l^2 \cos^2\theta}{r^2}
\nn
\\
\tilde{\Delta} &=& 1 + \frac{l^2}{r^2} +
\frac{2ml^2\sin^2\theta}{r^6 \Delta f}
\nn
\\
h&=& 1 - \frac{2m}{r^4\Delta}
\nn
\\
\tilde{h} &=& \frac{1}{\Delta}\left( 1+ \frac{l^2}{r^2} -
\frac{2m}{r^4} \right)
\dla{metricfns}
\ea
Note that this solution is not singular unlike the 4D Kerr
metric\footnote{The 4D Kerr metric can be obtained from the 10D metric in
the following way. First get rid of
six dimensions, namely $x^{i}$ and $\Omega_3$.  Then put
$\alpha=0$ (which is not too surprising since $\alpha$ contains
the string parameters).  Finally, for dimensional reasons, $m
\rightarrow m r^3$.}. The total
(quantized) D-brane charge of the
black hole is $R$ where $R^4 = 2m\sinh \alpha \cosh \alpha$.

The black hole rotates in the $\phi$-plane and its angular
momentum is determined by $l$: if $l=0$ the off-diagonal terms in
(\dr{rotD}) vanish and the metric is that of a D3 black brane
whose MC was considered in \dc{KK}.\footnote{We believe, however,
that there is a slight misprint in the Friedmann equation obtained
in \dc{KK} in that case: equation (5.6) of \dc{KK} should read $(E
+ \xi(1-a^4))^2/a^8$ rather than $(E + \xi a^4)^2/a^8$.} The
metric describing rotation about more than one axis was also
constructed in \dc{KLT}. Together with $\phi$, the coordinates $r$
and $\theta$ in (\dr{rotD}) are the usual coordinates describing a
3-sphere.  Notice that the metric coefficients are functions only
of $r$ and $\theta$ but not $\phi$. In a similar way to the
Sch-AdS$_5 \times$S$_5$ metric considered in the previous section,
there is a factorized 3-sphere contribution.

The horizon is the surface given by $\tilde{h}(r)
= 0$ so that
$$
r^2=r_0^2  = \ha \left( \sqrt{l^4 + 8m} - l^2 \right).
$$
A second critical surface, the infinite red-shift
hyperplane, satisfies
$h(r)=0$. We have
\be
r^2=r_{\infty}^2 = \ha \left( \sqrt{l^4\cos^4\theta + 8m} - l^2
\cos^2\theta \right)
\dle{rcdef}
\ee
and that $r_{\infty} \geq r_0$, with equality holding at the poles
($\theta =
0,\pi$).

It is convenient to introduce the function
\be
f_0 = 1 + \frac{R^4}{r^4 \Delta}.
\dle{f0def}
\ee
Then the 4-form potential for this background is
\dc{KLT}
\be
{C}_4  =  - \left( \frac{1 - f_0}{f} \right) - \frac{l \sqrt{2m}
\tilde{R}^2}{r^4 \Delta f} \sin^2 \theta \dot{\phi}.
%=
% -\xi  \left( \frac{1 - f}{f} \right) - \frac{l \sqrt{2m}
%\tilde{R}^2}{r^4 \Delta f} \sin^2 \theta \dot{\phi}
\dle{C4rot}
\ee

Finally, note that the brane scale factor is given by $a\!=\!f^{-1/4}$.
However, as we shall see
shortly, we consider trajectories in which $\theta\!=\!\pi/2$. Then
$\Delta\!=\!1$ so that
\be
a = \left(1 + \frac{\tilde{R}^4}{r^4}
\right)^{-1/4}.
\dle{Srot}
\ee
Thus $a$ is bounded by $a_{max}=1$ (as $r \rightarrow \infty$)
and $a_{\min} = (1 + \tilde{R}^4/r_0^4)^{-1/4}$ (as $r
\rightarrow r_0$).

\subsection{Brane dynamics with no matter}\dle{ss:dynrot}

In the static gauge, substitution of the metric (\dr{rotD}) into
the action (\dr{SDBIdef}) gives \dc{Cai,Alex}\footnote{In fact our
Lagrangian differs from theirs by an irrelevant overall constant.}
\be
{{\cal L}} = - \frac{1}{f} \left[ \sqrt{h-f\omega^2} - 1 + f_0 +
\frac{l \sin^2 \theta}{\sinh \alpha} (1-f) \dot{\phi} \right]
\dle{Lrot}
\ee
where we have set $q=1$ and
\be
\omega^2
% &=&   \left[\frac{1}{\sqrt{f} } \right] (2 g_{0\phi}
%\dot{\phi} + g_{rr} \dot{r}^2 + g_{\theta \theta} \dot{\theta}^2 +
%g_{\phi \phi} \dot{\phi}^2 + h_{IJ}\dot{\Phi}^I \dot{\Phi}^J )
%\nn
%\\
= \frac{\dot{r}^2}{\tilde{h}} - \frac{4 m l \cosh \alpha} {r^4
\Delta f} \sin^2 \theta \dot{\phi} + r^2 \left( \Delta
\dot{\theta}^2 + \tilde{\Delta} \sin^2 \theta \dot{\phi}^2 +
\cos^2\theta \dot{\Omega}_3^2 \right).
\dla{omegadef}
\ee
It is straightforward to obtain the equations of motion for
$\theta$ from (\dr{Lrot}) and to show that $\dot{\theta} =
\ddot{\theta} = 0$ if either $\sin \theta = 0$ or $\cos \theta =
0$. Hence the brane will remain in the same $\theta$-plane for
$\theta = 0,\pi/2$.

If $\theta=0$, then the coefficient of the $\dot{\phi}$ terms
vanish in (\dr{Lrot}) and (\dr{omegadef}) so that one is left with
diagonal metrics of the form discussed in (\dr{genmet}). Since
we want to study the effect of black hole rotation on brane
dynamics, we choose $\theta = \pi/2$ so that $r_{\infty}^2 = \sqrt{2m}$
and the Lagrangian becomes
\be
{\cal L} = - \sqrt{\cA + \cB \dot{r}^2 + \cC\dot{\phi}^2 + 2 \cD
\dot{\phi}}
 + \cE + \cG \dot{\phi}
 \equiv  -\sqrt{Z} + \cE + \cG \dot{\phi}
\dle{lsimple}
\ee
where
\ba
\cA  =  - g_d^3 g_{00} =  \frac{h}{f^2}, & \qquad & \cB  =  -
g_d^3 g_{rr} = - \frac{1}{f \tilde{h}},
\nn
\\
\cC  =  - g_d^3 g_{\phi \phi} = -\frac{r^2}{f}\tilde{\Delta},
&\qquad & \cD  =  - g_d^3 g_{0 \phi}
%\frac{2 m l \cosh \alpha} {r^4 \Delta
%f^2} \sin^2 \theta
=\frac{ l \sqrt{2m} }{r^4 f^2}\frac{R^4}{\tilde{R}^2},
\dla{AB}
\ea
and from (\dr{C4rot})
\ba
\cE & = &\frac{(1 - f_0)}{f} = - \frac{R^4}{r^4 f}
\nn
\\
\cG & = &\frac{l }{\sinh \alpha} \frac{(f-1)}{f} = \frac{l
\sqrt{2m} }{r^4  f} \tilde{R}^2.
\dla{C}
\ea
Notice that the coefficient of
$d\Omega_3^2$ in (\dr{rotD}) vanishes when $\theta = \pi/2$ so
that the brane can have no angular momentum about this 3-sphere;
hence this is a different set up from the one considered in the
previous section.  However, the brane does have a conserved
angular momentum about the $\phi$ direction: $\ell = \partial {\cal
L}/\partial \dot{\phi}$.

From (\dr{lsimple}) the angular momentum $\ell$ and energy $E$
are given by
\be
\ell = \cG - \frac{1}{Z^{1/2}} (\cC \dot{\phi} + \cD), \qquad E =
 \frac{1}{Z^{1/2}} (\cC \dot{\phi} + \cA) - \cE
%- \left[ \frac{1}{Z^{1/2}} (b \dot{r}^2 + c\dot{\phi}^2 + d
%\dot{\phi} ) - Z^{1/2} + e \right]
\ee
so that
\be
\dot{\phi} = -\frac{\cA \tilde{\cG} + \cD \tilde{\cE}}{\cM} \qquad
\dot{r}^2 = \frac{(\cA \cC-\cD^2)}{\cB} \frac{ [-\cA\tilde{\cG}^2
-\cC\tilde{\cE}^2 - 2\cD\tilde{\cG}\tilde{\cE} + (\cA \cC
-\cD^2)]}{\cM^2}
\dle{result}
\ee
where we have defined
\be
\tilde{\cG} \equiv -\ell+\cG, \qquad \tilde{\cE} \equiv -E-\cE, \qquad \cM
\equiv\cD\tilde{\cG} + \cC\tilde{\cE}.
\dle{tildedef}
\ee
Notice that on setting $\cD=\cG=0$ equations (\dr{result}) reduce
to (\dr{dotr}) as required. Finally, the brane time $\tau$ is
obtained by substituting (\dr{result}) into (\dr{eta}) to give
\be
d\tau^2 = \frac{1}{g_d^3} (\cA + \cB \dot{r}^2 + \cC
\dot{\phi}^2 +2\cD \dot{\phi}) dt^2 = \frac{1}{g_d^3 \cM^2}(\cA
\cC-\cD^2)^2 dt^2.
\dle{etarot}
\ee

We are now in a position to construct the different effective
potentials defined in section \dr{ss:potSch}. The first, $V_{\bf
eff}^t(r,\ell,E)$, is given by
\be
V_{\bf eff}^t(r,\ell,E) = E -
 \frac{(\cA \cC-\cD^2)}{2\cB} \frac{ [-\cA\tilde{\cG}^2 -\cC\tilde{\cE}^2
- 2\cD\tilde{\cG}\tilde{\cE} + (\cA \cC-\cD^2)]}{\cM^2}. \dle{Vtrot}
\ee
In \dc{Cai}, an attempt was made to study this potential in the limit that
$\dot{\phi} =0 = \ell$.  However, it is clear from (\dr{result}) that this
is not a consistent choice: if $\ell = 0$ then $\dot{\phi}=0$ only for a
very specific value of $r$, namely when $\cA \cG = \cD (E+\cE)$. Instead,
one should study (\dr{etarot}) for arbitrary $\ell$.

First notice that at the horizon $r\!=\!r_0$, $\cB^{-1}=0$ so that
$V_{\bf eff}^t(r_0) = E$ and $\left. \partial V_{\bf
eff}^t/\partial r  \right|_{r_0} = 0$. This is just as for the effective
potential discussed in section \dr{ss:potSch}. Now, however, as $r
\rightarrow \infty$,
\be
V_{\bf eff}^t(r) \rightarrow E + \frac{1}{2E^2}
( 1 - E^2)
\dle{Vrot}
\ee
so that only for $|E|>1$ will the brane be able to escape from the
rotating black hole: whenever $|E|<1$ the brane is trapped.
The behaviour of $V_{\bf eff}^t$ as a function of $\ell$---the
brane angular momentum---is similar to that discussed in
section \dr{ss:potSch}. For all values of $l$ and $m$, there is a
critical value of the angular momentum $\ell_c$ above which a
repulsive centrifugal barrier forms.  At $\ell = \ell_c$ (where
$\ell_c$ is of course a function of $l,m$ and the other
parameters) there is a corresponding critical radius $r=r_c$ in
which the brane is in a stable circular brane orbit with constant
angular velocity $\dot{\phi}_c$. (For any other value of $\ell$,
the radius $r$ is not constant and hence, from (\dr{result}),
$\dot{\phi}$ is not constant either.)

Given the equations of motion above, one may ask if there is a
solution in which the the relative position of the brane to the
rotating source is constant. In other words, is there a solution
$\dot{\phi}_c = \Omega$ where $\Omega$ is the angular velocity of
the black hole given by $\Omega = \sqrt{2} R^{-2} m^{-1/2} l
r_0^2$ \dc{KLT,Cai2}?  In reference \dc{Cai} it was assumed that
such a solution exists and the thermodynamics of the D3-brane was then
studied as a function of $r$ (since for a static
probe, its distance to the source can be regarded as a mass scale
in the SYM theory \dc{TY}). In particular, by calculating the
entropy and heat capacity of the brane for $\theta = \pi/2$ and
$\dot{\phi}_c = \Omega$, it was shown that there are two critical
points for which these thermodynamic quantities diverge leading to
interesting conclusions regarding the mass scale of scalar fields
in SYM theory \dc{Cai}.  Our analysis of (\dr{Vrot}) suggests,
however, that generically $\dot{\phi}_c \neq \Omega$.  For a given
set of $(l,m,\alpha)$, the value of $\dot{\phi}_c$ depends on $E$
and there is only one specific value of $E \equiv E_c$ for which
$\dot{\phi}_c = \Omega$.  If for some reason these specific values
of $(\ell_c,E_c)$ are chosen (this is a set of measure zero) then
the radial distance of the brane, $r=r_c$, is also fixed. Hence it
does not appear consistent with equations (\dr{result}) to study
probe brane thermodynamics by setting $\dot{\phi}_c = \Omega$ and
then letting $r$ vary.

We now turn to the cosmologically relevant effective potential
$V_{\bf eff}^\tau(r,\ell,E)$ given by
\be
V_{\bf eff}^\tau(r,\ell,E) = E - \ha \frac{g_d^3}{\cB(\cC
\cA- \cD^2)} \left[-\cA\tilde{\cG}^2 -\cC\tilde{\cE}^2 -
2\cD\tilde{\cG}\tilde{\cE} + (\cC \cA-\cD^2) \right].
\ee
In the limit $r \rightarrow \infty$,
\be
V_{\bf eff}^\tau \rightarrow E - \ha (E^2-1)
\dle{Vrotfinal}
\ee
so that once again for $|E| < 1$ the brane cannot escape from
the rotating black hole. Notice that there is a significant difference
between $V_{\bf eff}^\tau$ in this rotating black hole bulk
and that obtained for the Sch-AdS$_5 \times$S$_5$ bulk: there
the brane (with $q=1$) always had zero
kinetic energy at infinity since  $V_{\bf eff}^\tau(r\rightarrow \infty)
=E$.  In other words the cosmological constant on the brane vanished. Here, on
the other hand, it is clear from (\dr{Vrotfinal}) that even when
$q=1$, the cosmological constant only vanishes when $E=1$ (as in that
case the brane has no kinetic energy at infinity).  This is the first
indication that the dark fluid terms in the Friedmann equation
will be rather different in this rotating black hole background (see
section \dr{ss:Friedrot}).

In the limit $r \rightarrow r_0$, $V_{\bf eff}^\tau < E$, and notice also
that the coefficient of the $\ell^2$ term is positive so that once again
we expect a centrifugal potential barrier. The brane motion can be
summarised as follows
\begin{itemize}
\item $E < 1$.  Independently of
$\ell$, the brane will be trapped in a region near the horizon and
eventually be absorbed by the black hole.
\item $E \geq 1$.  Here one has
similar behaviour to the one in the Sch-AdS$_5 \times$S$_5$ background.
That is, there is again a critical angular momentum $\ell_c$ which divides
the brane trajectories into two different categories as discussed in
section \dr{ss:dynSch}.
\end{itemize}
The behaviours are illustrated in figure \dr{fig:Veffroteta} for
$E<1$ and figure \dr{fig:VeffrotetaE} for $E>1$, where we have introduced
the
dimensionless quantities $\hat{r}\!=\!r/r_{\infty}$,
$\hat{l}\!=\!l/r_{\infty}$,
$\hat{R}\!=\!R/r_{\infty}$ and $\hat{\ell}\!=\!\ell/r_{\infty}$.
\begin{figure}[ht]
\begin{center}
\psfrag{r}[b]{\hspace{9.3cm}$\sqrt{2}\hat{r}$}
\psfrag{6}[]{\hspace{-0.2cm}$V_{\bf eff}^{\tau}$}
\psfrag{2.5}[]{}
\psfrag{3.5}[]{}
\psfrag{4.5}[]{}
\includegraphics[]{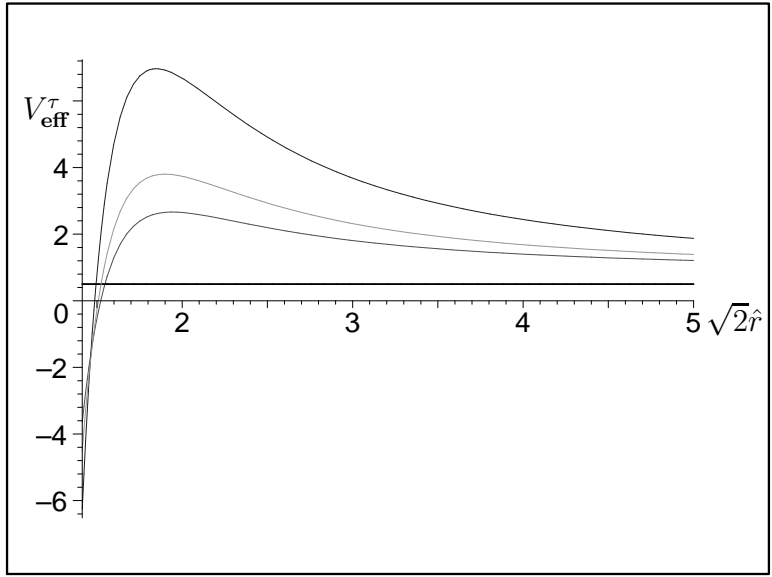}
\caption{The
effective potential $V^\tau_{\bf eff}$.  The parameters are $q=
1$, $\hat{l}^2=0.5,\hat{R}^4=4$ (so that $r_0 = 0.88\,r_{\infty}$), the
angular
momentum
$\hat{\ell}^2=0$ and the energy $E=0.5$. As discussed in the text, the
brane will inevitably be trapped and fall into the black hole.}
\label{fig:Veffroteta}
\end{center}
\end{figure}

\begin{figure}[ht]
\begin{center}
\psfrag{r}[b]{\hspace{9.3cm}$\sqrt{2}\hat{r}$}
\psfrag{0}[]{\hspace{-0.2cm}$\hat{V}_{\bf eff}^{\tau}$}
\psfrag{2.5}[]{}
\psfrag{3.5}[]{}
\psfrag{4.5}[]{}
\includegraphics[]{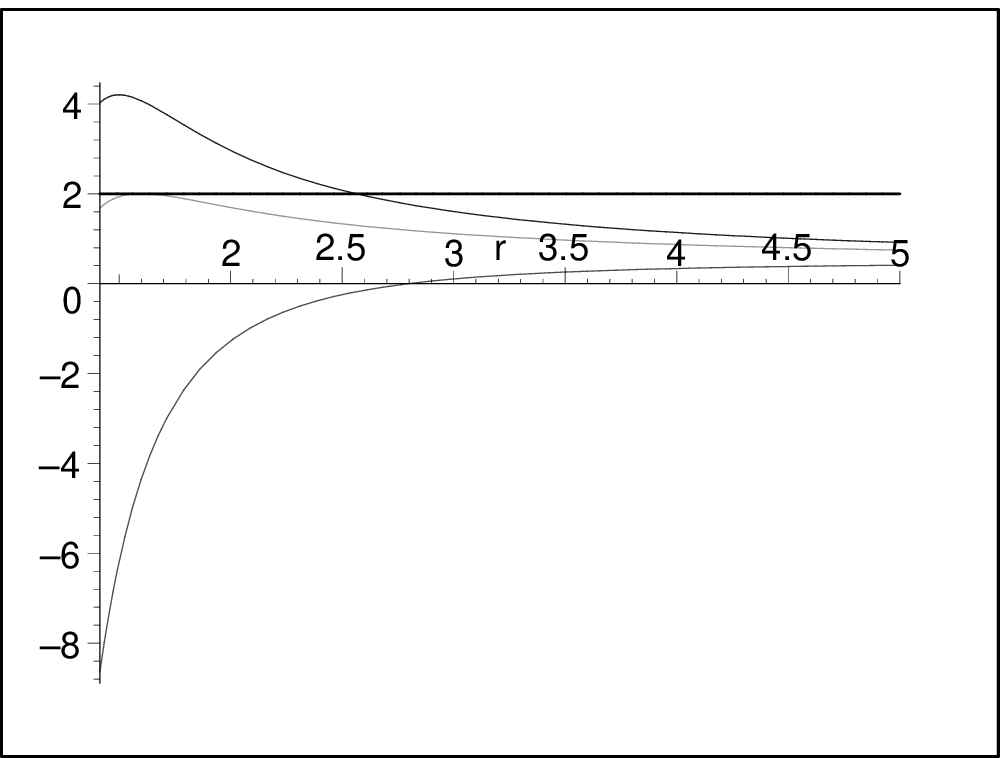}
\caption{The
effective potential $V^\tau_{\bf eff}$.  The parameters are $q=
1$, $\hat{l}^2=0.5,\hat{R}^4=4$, and the energy $E=2$. Once again there
are the
three possible regimes, depending on the angular momentum. Here
the values are $\hat{\ell}^2=0$ for the lower curve,
$\hat{\ell}^2=\hat{\ell}^2_c=2.038$
for the critical curve, and $\hat{\ell}^2=2.5$ for the upper curve.}
\label{fig:VeffrotetaE}
\end{center}
\end{figure}

In this rotating black hole background, the existence of stable
circular orbit with $r>r_0$ was required by Alexander \dc{Alex}
who studied the effects of a varying speed of light in MC (see
also section \dr{ss:caus}).  In \dc{Alex}, however, a rather
involved mechanism was constructed to stabilise the brane in a
circular orbit at some $r> r_0$  (this was required since the only
stable circular orbit was thought to have radius $r=r_0$ where the
speed of light vanishes (see section \dr{ss:caus})). According to
our analysis, however, a stable circular orbit with $r>r_0$ always
exists when $\ell = \ell_c$---and given the analysis of section
\dr{s:Sch} this is indeed the case whether or not the black hole
rotates.

\subsection{Friedmann equation}\dle{ss:Friedrot}

We end this section with a few comments regarding the dark fluid
terms which appear in the Friedmann equation when doing MC in this
rotating black hole background.  As mentioned above, we expect an
$E$-dependent cosmological constant.  Combination of equations
(\dr{etarot}),(\dr{Fried}) and (\dr{result}) yield the following
Friedmann equation:
\be
H^2
%\left( \frac{1}{S} \frac{d S}{d\tau} \right)^2 &=&
%\frac{ \dot{r}^2}{4}
%\left(\frac{dt}{d\tau} \right)^2 \left( \frac{g_D'}{g_D} \right)^2
%\nn
%\\
%& = &
= \frac{g_d (g_d')^2}{4\cB(\cC \cA -\cD^2)}  \left[-\cA
\tilde{\cG}^2 -\cC \tilde{\cE}^2 - 2\cD \tilde{\cG}\tilde{\cE} +
(\cC \cA -\cD^2) \right].
\dle{H2Rot}
\ee
Unfortunately,
the right-hand side of this equation cannot simply be written as a
sum of terms of the form $1/a^p$ for some power $p$, since now $r$
is not a simple function of the scale factor $a$: inversion of
(\dr{Srot}) yields
\be
{r} = {\tilde{R}} \frac{a}{(1-a^4)^{1/4}}.
\dle{rS}
\ee
Thus it is not possible simply to read off the dark fluid terms; we
conclude that these terms are background dependent
(the same conclusion was reached in \dc{KK} as a result of
studying a number of different static backgrounds). However, one
can study the behaviour of $H^2$ when $a\ll 1$.
Then $r \simeq \tilde{R}a$, and a straightforward Taylor expansion yields
\be
H^2 \simeq \frac{c_0 + c_2 a^2 + c_4 a^4 + c_6 a^6}{\tilde{R}^6
a^{10}}
\ee
where
\ba
c_6 & =& \tilde{R}^6(E^2-1)
\nn
\\
c_4 & = & \tilde{r}^4 \left( l^2E^2 - \ell^2 -
\frac{l^2}{\tilde{R}^2} \right)
\nn
\\
c_2 & = & \tilde{R}^4 (\xi + E)^2
\nn
\\
c_0 & = & \tilde{R}^2 \left( l^2(1+E\xi)^2 + \ell^2(\xi^2-1)
\right) - 2 l \ell \sqrt{2m} (1+\xi E)
\nn
\ea
and
$$
\xi = \frac{R^4}{\tilde{R}^4} = \frac{\cosh \alpha}{\sinh \alpha}
= \sqrt{1 + \frac{2m}{\tilde{R}^4}}.
$$
Hence as in the case of the Sch-AdS$_5 \times$S$_5$ black hole of
section \dr{s:Sch}, for small $a$, $H^2$ contains dark fluid
terms proportional to $a^{-10}$, $a^{-8}$, $a^{-6}$ and $a^{-4}$.
However, it is important to notice now that even if the angular
momentum $\ell$ of the brane vanishes, there are still the
contributions going as $a^{-6}$ and $a^{-10}$ in the Friedman
equation.  These are now sourced by the angular momentum ($\propto
l$) of the black hole itself rather than that of the brane.

\section{Comments on causality and brane inflation}
\dle{s:varyc}

\subsection{Causality}\dle{ss:caus}

In brane world scenarios it is well known that Lorentz invariance
is violated \dc{Lang}.  This reflects the fact that gravitons can
propagate in the bulk whereas photons are confined to the brane:
hence gravitational and light signals generally take different
times to propagate between two given points on the brane.  In the
context of MC, varying speed of light effects have been discussed
by \dc{Kehag2,Alex} for slowly moving branes (we will be more
specific about the meaning of `slowly moving' below).  Here we
comment briefly on this varying speed of light, $c_{\bf eff}$,
without making any approximation regarding the brane dynamics
since this was obtained exactly in sections \dr{ss:dynSch} and
\dr{ss:dynrot}.  (In fact it is not entirely clear to us whether
this `varying speed of light' effect should not be referred to as a
redshift effect.  However we use the terminology `varying speed of
light' as in references \dc{Kehag2,Alex}.)

In references \dc{Joao} an investigation was made of the
cosmological problems which may be resolved if $c_{\bf eff}$ was
always larger in the past.  Thus in MC we search for an expanding
universe for which $c_{\bf eff}$ is always a decreasing function
stabilising at a constant value $c_0$ as $\tau \rightarrow
\infty$.  Notice that in \dc{Alex} the universe-brane was always
considered to approach the black hole (corresponding to a
contracting universes).  We consider the case when the brane moves
radially outwards and hence expands: how does $c_{\bf eff}$ behave
in that case?

When photons are present on the brane $F_{ij} \neq 0$ (we still
keep $B_{ij}=0$).  Expansion in powers of $\alpha' \ll 1$ of the
D-brane action (\dr{action}) in the static gauge for $q=1$ yields,
to second order,
\ba
S &\simeq& - \lambda \int d^4 x \sqrt { - \det \gamma_{ij}
 } - \lambda \int d^4 x \hat{C}_4
 \nn
 \\
 & + &  (2 \pi \alpha')^2 \;
 \frac{\lambda}{4}\int
d^4 x \sqrt { - \det \gamma_{ij} }\; \times  \tr \left[
\bgamma^{-1} \; \matr{F}
 \bgamma^{-1} \; \matr{F}\right] + \ldots.
\dla{DBId}
\ea
Here $\matr{A} \equiv A_{ij}$, and note that the term linear in
$\alpha'$ gives no contribution since $\matr{F}$ is anti-symmetric
so that $\tr \; {{\bgamma}^{-1} \; \matr{F}} = 0$.

Since $\alpha' \ll 1$ to first order the brane dynamics will be
governed by the first two terms of (\dr{DBId}), and hence will be
given by (\dr{dotr}) for a Sch-AdS$_5 \times$S$_5$ background, or
by (\dr{result}) for the rotating black hole background.
Furthermore
\be
 \tr \left[ {\bgamma}^{-1} \; \matr{F}
{\bgamma}^{-1} \; \matr{F} \right] =  \gamma_d^{-1} \left[
 -\gamma^{00} {\bf E}^2 + \gamma_d^{-1} {\bf B}^2 \right]
\ee
where $F_{0i} = E_i$ and $F_{ij} = \epsilon^{ijk} B_k$ and we have
defined $\gamma_{ab} \equiv \gamma_d \delta_{ab}$. Hence (\dr{DBId}) is a
kinetic term
for the gauge fields:
\be
(2 \pi \alpha')^2 \frac{\lambda}{4}\int d^4 x \sqrt { - \det
\gamma_{ij} } \; \times \tr \left[ {{\bgamma}^{-1} \; \matr{F}
 {\bgamma}^{-1}} \matr{F} \right] \; = \;
(2 \pi \alpha')^2 \; \frac{\lambda}{2}  \int d^4 x \left( -
\tilde{A} {\bf E}^2  + \tilde{B} {\bf B}^2 \right)
\ee
where
\be
\tilde{A} = \left( \frac{\gamma_d}{|\gamma_{00}|} \right)^{1/2}
\qquad \tilde{B} = \left( \frac{|\gamma_{00}|}{\gamma_d}
\right)^{1/2},
\ee
so that the effective speed of light is
\ba
c_{\bf eff} & =  & \left( \frac{|\gamma_{00}|}{\gamma_d}
\right)^{1/2}
\nn
\\
&=& \frac{- {\cal L}(X) - q C_4(X) }{g_d^2}.
\dla{ceff}
\ea
Here we have used the definition of ${\cal L}$ given in
(\dr{Lagdef}). Notice that ${\cal L}$ and $C_4$ must be evaluated
on the brane trajectory $X(\tau)$ so that the effective speed of
light clearly depends on the dynamics of the brane itself.

Since we search for a scenario in which $c_{\bf eff}$ tends
asymptotically to $c_0$ as the universe expands, there are two
cases to consider: {\it i)} either $\ell < \ell_c$ so that the
universe expands reaching $r \rightarrow \infty$ as $\tau
\rightarrow \infty$, or {\it ii)} $\ell = \ell_c$ in which case
the scale factor stabilises at a finite value of $r=r_c$.  We
analyse these cases first for the non-rotating
Sch-AdS$_5\times$S$_5$ background of section \dr{s:Sch}.

\subsubsection{Varying speed of light in Sch-AdS$_5\times$S$_5$}

In this case equation (\dr{ceff}) becomes
\be
c_{\bf eff} = \frac{1}{g_d^2} \sqrt{ \left( \cA + \cB \dot{r}^2 +
\cC h_{IJ} \dot{\phi}^{I} \dot{\phi}^{J} \right)}.
\dle{cnonrot}
\ee
If the brane moves slowly \dc{Alex}, then the first term in the
square-root dominates and $c_{\bf eff} \simeq \sqrt{\cA}/g_d^2 =
\sqrt{f(r)} = \sqrt{1-(r_0/r)^4}$ (where we have used (\dr{AdS})
and (\dr{abcedef})). This result is independent of $E$ and $\ell$
(respectively the energy and angular momentum of the brane) and of
$L$. Also $c_{\bf eff} = 0$ at $r=r_0$ and $c_{\bf eff}
\rightarrow 1$ as $r\rightarrow \infty$ so that $c_{\bf eff}$
increases as the universe expands.

If the brane does not move slowly, substitution into
(\dr{cnonrot}) of the specific expressions for $\dot{r}^2$ and
$h_{IJ} \dot{\phi}^{I} \dot{\phi}^{J}$ given in (\dr{dotr}) yield
\be
c_{\bf eff} = \frac{1}{g_d^2} \frac{\cA}{E + \cE} = \left[ 1-
\left(\frac{r_0}{r}\right)^4 \right] \left(
\frac{\tilde{E}L^4}{r^4} + 1 \right)^{-1}.
\dle{ceffnonrot}
\ee
Hence $c_{\bf eff}$ now depends on the energy $E$ of the brane,
though not on its angular momentum $\ell$. However, we still have
that $c_{\bf eff} = 0$ at $r=r_0$ and that $c_{\bf eff}
\rightarrow 1$ as $r\rightarrow \infty$.  Also, from
(\dr{ceffnonrot}), $c_{\bf eff}$ is a strictly increasing function
of $r$. Figure \dr{f:ceffnonrot} shows $c_{\bf eff}$ and the
effective potential $V^{\tau}_{\bf eff}$ when $\ell = \ell_c$.
\begin{figure}[ht]
\begin{center}
\psfrag{r}[b]{\hspace{9.3cm}$\hat{r}$}
\psfrag{1.5}[]{\hspace{2.8cm}$\hat{V}_{\bf eff}^{\tau}$}
\psfrag{0.5}[]{\hspace{3.3cm}$c_{\bf eff}$}
\psfrag{-0.5}[]{}
\includegraphics[]{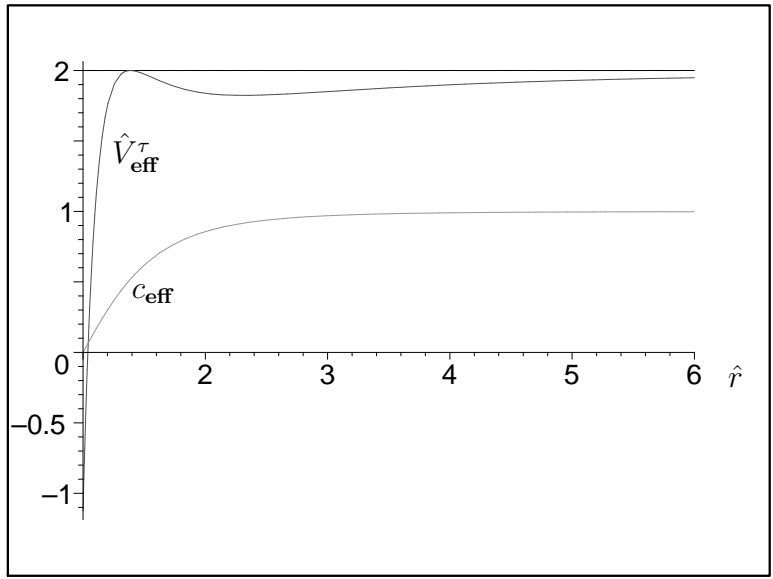}
\caption{The rescaled
effective potential $V^\tau_{\bf eff}$ for $\hat{\ell} = \hat{\ell}_c$
(upper
curve) and effective speed of light $c_{\bf eff}$ (lower curve) in
a Sch-AdS$_5 \times$S$_5$ bulk.  The parameters are as in figure 3
so that $q=1$, $\hat{L} =1$ and $\hat{E}=2$.  The upper horizontal line is
the energy $\hat{E}=2$.} \label{f:ceffnonrot}
\end{center}
\end{figure}
Thus once again, as the brane moves outwards from the horizon and
expands, $c_{\bf eff}$ decreases.  Furthermore, if $\ell = \ell_c$
then $c_{\bf eff}$ stabilises at a value near $1/2$. Hence the
only way in which $c_{\bf eff}$ can decrease and stabilise at late
times is if $\ell=\ell_c$ and the universe {\it contracts}.

\subsubsection{Varying speed of light in the rotating black hole
background}

A similar analysis for the bulk of section \dr{s:rot} yields
$$
c_{\bf eff} = \frac{1}{g_d^2} \frac{\cA \cC - \cD^2}{\cM}.
$$
Here $\cA, \cC, \cD$ are given in (\dr{AB}), $\cM$ in
(\dr{tildedef}) and we have chosen $\theta = \pi/2$ as in section
\dr{ss:dynrot}.  Notice that $c_{\bf eff}$ is now a function of
both $E$ and $\ell$.

In the limit $r \rightarrow \infty$, $c_{\bf eff} \rightarrow 1/E$
(this reflects the $E$-dependent cosmological constant in this
case).  One can also show that $c_{\bf eff}$ increases as $r$
increases.  Hence $c_{\bf eff}$ tends asymptotically to 1 as $r
\rightarrow \infty$ only if $E=1$. For this rotating black hole
background we do not present plots of $c_{\bf eff}$ corresponding
to the different curves in figure \dr{fig:VeffrotetaE}; the overall
behaviour of $c_{\bf eff}$ is similar
to that shown in figure \dr{f:ceffnonrot}.  For the parameters of
figure \dr{fig:VeffrotetaE} with $\ell = \ell_c$, $c_{\bf eff}\sim
0.2$ at $r=r_c$ and then tends asymptotically to $1/2$ as $r
\rightarrow \infty$.  Thus once again, the only way in which
$c_{\bf eff}$ can decrease and stabilise at late times is if $E
\geq 1$, $\ell=\ell_c$ and the universe contracts.

\subsection{Comments on non-relativistic brane matter and
inflation}\dle{ss:infln}

So far only the effects of radiation on the brane have been
considered in MC.  (To the best of our knowledge this is also true
of the rest of the literature on MC.)  How can non-relativistic
matter be included starting from the D-brane action?  The answer
probably lies in the fermionic sector of the string action which
has not been considered here.

As an alternative, one can take a more phenomenological approach
and add by hand matter on the brane with an arbitrary equation of
state \dc{PS}.  A byproduct of this approach is that, depending on
the bulk, it is possible to show that inflation can occur on the
brane (i.e.\ $a \sim \tau^{\alpha}$ with $\alpha
> 1$) \dc{PS}, but not in the bulk. An interesting
realisation of this occurs in the
following case: suppose the bulk is generated by a brane gas
\dc{Easson}, and consider the late time behaviour of this gas.  The
bulk metric, which is assumed to be flat and
roughly homogeneous and isotropic, is described by a
scale
factor $a(t)$ and depends on the dilaton field $\phi(t)$ and bulk
matter parameters $\rho$ and $p$
(energy density and pressure respectively). With the standard
embedding, a 3-brane moving in the bulk sees an induced scale factor
$a(t(\tau))$, where $\tau$ is the brane time. It is the difference between
$\tau$ and $t$ which is responsible for the different evolutions of the
brane and the bulk.

In this scenario, one considers a phenomenological
brane action of the form
$$
S = \int d^4 x \sqrt{-\gamma} {\cal L} = \int d^4 x \sqrt{-\gamma}
\left\{ e^{-\phi}\lambda + \xi e^{-m \phi} {\cal L}_b \right\}
$$
rather than (\dr{action}), where $m$ and $\xi$ are dimensionless constants
which determine the coupling of the dilaton to the brane matter
${\cal L}_b$.  Notice that any coupling to a 4-form has been
neglected. The first term above is just the general expression for
the kinetic term of (\dr{action}) for non-zero dilaton.  Given
this action it is not hard to solve for the brane dynamics, and
hence to obtain the brane scale factor in the way outlined in
section \dr{s:mirage}. Furthermore, if the brane initially has a
{\em large} velocity (for example, it is formed as the result of a
collision process
--- say a $\bar{5}-5$ brane annihilation \dc{bar}) and if $\rho_b
\gg \lambda$ then inflation may occur on the brane in the
radiation dominated epoch: in \dc{PS} this setup is analysed in
detail. The result is that for the natural coupling to the
dilaton, $m=1$, the brane inflates when the bulk is comprised of stiff
matter.

A rather different realisation of inflation in MC comes from
observing that if the brane moves slowly then the action
(\dr{action}) may be expanded in powers of $\dot{r}^2$ leading, to
first order, to action quadratic in $\dot{r}^2$. The field $r$ can
then be identified with the inflaton which now has an unusual
kinetic term, and combined with the potential term it can lead to
inflation---this set-up is reminiscent of that of
Burgess et al \dc{Quevedo}.  (For a discussion of this approach in
a bulk in which supersymmetry is broken, and the effects of brane 
self-gravity are included, see \dc{BrS}.)

\section{Conclusions}\dle{s:conc}

In this work we have tried to summarise some aspects of the approach to
brane cosmology known as mirage cosmology \dc{KK}. Here the brane is a
D3-brane in type IIB string theory and it moves in a 10D bulk metric. As
opposed to the 5D junction condition approach to brane cosmology, the
D3-brane is treated as a test brane and hence it is straightforward to
consider more than one extra dimension.

As explained in section \dr{s:mirage}, brane motion can induce an
effective cosmology on the brane, and once the dynamics of the
brane is determined the corresponding Friedmann equation can be
obtained.  Those parts of the Friedmann equation solely generated
by the motion of the brane (and not by matter on the brane) are
the dark fluid terms.  We have tried to see how the familiar dark
radiation term \dc{Binetruy} generalises when there is more than
one extra dimension, and this was done for the two specific 10D
bulk metrics of sections \dr{s:Sch} and \dr{s:rot}.

In section \dr{s:Sch} we studied the dynamics of the probe
D3-brane in a Sch-AdS$_5 \times$S$_5$ bulk for which the brane
geodesics are parametrised by a conserved energy $E$ and an
angular momentum $\ell$ about the $S_5$.  For all $\ell$ we saw
that the cosmological constant on the brane vanished if $q \equiv
e/\lambda = \pm 1$ corresponding to BPS (anti-)branes.  Also the
Friedmann equation was found to contain dark fluid terms
proportional to $a^{-4}, a^{-8}$ and to $\ell^2 a^{-6}, \ell^2
a^{-10}$.

When $\ell=0$, the brane motion is constrained to Sch-AdS$_5$.
There, however, the exact brane dynamics (including the
back-reaction of the brane on the bulk) can be calculated
\dc{Kraus,Ida,CU}. As discussed in section \dr{ss:linkMCJC}, the
MC results must be compared to a JC calculation in which $Z_2$
symmetry is broken since the D3-branes couple to the bulk RR
field. We saw that the MC and JC Friedmann equations had the same
dark fluid terms, and a further analysis of those equations linked
the parameters of the MC approach ($E$ etc) to those of the JC
approach (equations (\dr{resa})-(\dr{resss})). For that analysis
it was important to allow the D3-branes to have an arbitrary RR
charge $q$.

Non-zero angular momentum, $\ell \neq 0$, generated dark fluid
terms $\propto a^{-10}, a^{6}$.  As a result different types of
brane trajectories were seen to exist depending on whether or not
$\ell$ was greater or smaller than $\ell_c$ (figure
\dr{fig:Veffeta}):
\begin{itemize}
\item $\ell < \ell_c$.  The brane contracts/expands (corresponding
to inward/outward radial motion) for all $\tau$.
\item $\ell = \ell_c$.  A (contracting) brane moving radially inwards from
infinity reaches, after an infinite $\tau$, a critical radius in
which it rotates around the black hole in a stable orbit.  An
expanding brane moving radially outwards from the horizon reaches
the same stable critical radius.
\item $\ell > \ell_c$.  A centrifugal barrier develops.  A (contracting)
brane moving radially inwards from infinity bounces off this barrier
after a finite $\tau$. Then it moves radially outwards and starts
to expand. Similarly a (expanding) brane moving radially outwards
from the horizon is also reflected by the barrier after a finite
$\tau$; it starts moving radially inwards and contracts until it is
swallowed by the black hole.
\end{itemize}

In section \dr{ss:radSch} we commented on the addition of radiation to the
brane.  The resulting Friedmann equation only contains terms proportional
to $\rho_{rad}$ and not $\rho_{rad}^2$ because of the `passive' nature of
the MC approach.  In principle, given this Friedmann equation, one could
try to constrain the different parameters (such as $\ell$, $E$) via
nucleosynthesis constraints.  One reason for not doing this is the
(current) lack of a treatment for non-relativistic brane matter in this MC
approach. Since D3-branes are BPS states, perhaps this problem will be
related to providing a prescription for supersymmetry-breaking. How to do
all this is an important question for future work.

The purpose of section \dr{s:rot} was to consider a slightly more
complicated bulk and to try to see how many of the results
presented in section \dr{s:Sch} are in fact bulk dependent.  We
considered the dynamics of a brane in the rotating black hole
metric of equation (\dr{rotD}) and found that the main differences
with the Sch-AdS$_5 \times$S$_5$ bulk are
\begin{itemize}
\item an $E$-dependent cosmological constant on the brane which
does not vanish when $q=\pm 1$ unless $E=1$.
\item Brane trajectories which were always trapped by the
black hole if $E<1$.  For $E>1$ the three different classes of
$\ell$-dependent trajectories outlined above were found.
\item  For $a \ll 1$, dark fluid terms $\propto a^{-10}, a^{-6}, a^{-8}$ and
$a^{-4}$, the first two of which did {\em not} vanish when
$\ell=0$ since in this case they were sourced by the angular
momentum of the black hole itself (i.e.\ $\propto l$).
\end{itemize}

Finally for both bulks, we considered the behaviour of $c_{\bf
eff}$ as the universe expands.  In the Sch-AdS$_5 \times$S$_5$
bulk $c_{\bf eff}$ is $\ell$ independent (but $E$ dependent) and
vanishes at the horizon $r_0$.  As $r \rightarrow \infty$, $c_{\bf
eff}$ tends to 1. Hence for a brane with $\ell < \ell_c$ (which
expands for all $\tau$), the speed of light always increases
tending to 1.  For $\ell = \ell_c$ the asymptotic value is $c_0 <
1$.  Similar behaviour holds in the rotating black hole bulk.

There are many interesting aspects of mirage cosmology which we
have not studied here.  One of these is the question of the
initial singularity \dc{KK}, and another is an interpretation of
the results presented here (and especially the role of the
critical angular momentum $\ell_c$) in the context of SYM theory
and black hole thermodynamics.

A final crucial ingredient, which is required for MC, is a description of
brane self-gravity. As was discussed in the introduction, our approach is
to treat brane motion as similar to planetary motion. In the latter case,
one usually leaves the question of self-gravity to the geophysicists.
However, as cosmologists, it is necessary to know about the internal
evolution of the brane. Since it is extremely unlikely this will be
induced solely by the motion of the brane in its background, we require an
understanding of how local energy density on the brane sources gravity on
the brane \dc{BrS}. Before mirage cosmology can become a fully-fledged 
cosmology,
this vital question must be addressed.

\section*{Acknowledgements}

D.A.S.\ would like to thank the organisers and participants of
Peyresq-6 for a very enjoyable and interesting meeting. We thank
S.~Alexander, T.~Boehm, Ph.~Brax, R.~Durrer, E.~Kiritsis,
M.~Ruiz-Altaba and J-Ph.~Uzan for useful discussions or comments.
Our thanks to the Theoretical Physics Groups of both Geneva
University and Imperial College, London, where the majority of
this work was done. Finally we would like to thank Nick Rivier and
Mme.\ Rivier-Mercier for wonderful hospitality in Verbier,
Switzerland while this work was in progress.

\section*{Appendix}

In section \dr{ss:potSch} we commented that there is a critical value of
the brane angular momentum, $\ell_c$, for which the effective potential
has a local maximum at $V_{\bf eff}\!=\!E$. A brane with this angular
momentum can be in a circular orbit about the black hole at $r\!=\!r_c$.

Mathematically, this situation arises when $V_{\bf eff}\!=\!E$
has repeated roots. If we use the rescaled quantities and let
$x\!=\!\hat{r}^2$, then from (\dr{Vefft}) or (\dr{Veffeta}) this
is equivalent to considering when the polynomial
$p(x)\!=\!(x^2-1)(\hat{\ell}^2+x^3)-x(\hat{E}-\ha+x^2)^2\!=\!0$.
Simplifying, it is left to show that there are repeated roots of
the cubic equation
$$ y x^3 - \hat{\ell}^2 x^2 + \qa(y-1)^2 x +
\hat{\ell}^2 =0, $$
where $y=2\hat{E}$. This occurs when
$$
\hat{\ell}^4_c = \frac {1}{128}\left[
{-y^4+76y^3+282y^2+76y-1+\sqrt{(y+1)^2 (y^2+34y+1)^3}}\right].
$$
The expression for the repeated root $x_c\!=\!\hat{r}_c^2$ is too
complicated to write down here however.

\typeout{--- No new page for bibliography ---}

\end{document}